\documentclass[11pt,a4paper]{article}

\usepackage[margin=1in]{geometry}
\usepackage{booktabs}
\usepackage{tabularx}
\usepackage{multirow}
\usepackage{makecell}
\usepackage{array}
\usepackage{xcolor}
\usepackage[hidelinks]{hyperref}
\usepackage{graphicx}
\DeclareGraphicsExtensions{.jpg,.jpeg,.png}
\usepackage{natbib}
\usepackage{authblk}
\usepackage{abstract}
\usepackage{titlesec}
\usepackage{parskip}
\usepackage{amsmath}
\usepackage[T1]{fontenc}

\title{\textbf{Broken Object Level Authorization in the Wild: An Empirical
  Taxonomy from 100+ Bug Bounty Disclosures}}

\author[1]{Bandana Kaur}
\affil[1]{APIsec Research Labs \\ \texttt{bandana@hackwither.co.in}}   

\date{}  

\begin{document}
\maketitle

\begin{abstract}
Broken Object Level Authorization (BOLA) is consistently ranked the most critical API security vulnerability, yet the existing literature remains almost entirely conceptual. This paper presents one of the first large-scale empirical analyses of BOLA in publicly disclosed bug bounty reports. We constructed a reproducible sampling frame of 200 HackerOne disclosures tagged IDOR or Improper Access Control (2021–2026) and applied a three-criterion inclusion filter, yielding 107 fully classified reports. Classification used an LLM-assisted schema-completion procedure under constrained, human-adjudicated criteria against a six-family BOLA taxonomy. Of 107 classified reports, 84 (78.5\%) were confirmed in-scope BOLA. Action-Level Object BOLA, defined by unauthorized state-changing actions on another user's objects, accounts for 41.7\% of confirmed cases and emerges alongside Direct Object Reference BOLA as one of the two dominant families observed in the dataset. This shows a pattern historically underrepresented in practitioner guidance. Approximately 21.5\% of classified reports are out-of-scope under strict criteria, indicating that tag-counting on platforms like HackerOne significantly overstates the BOLA-specific signal. We report distributions across family, action type, authorization direction, industry sector, identifier format, and exploit mechanism. Key secondary findings include an 11.9\% rate of vertical (user-to-admin) privilege failures and systematic exploitation of GraphQL Global IDs across major platforms. Findings have direct implications for API security testing protocols, developer education, and OWASP guidance.

\end{abstract}

\section{Introduction}

Broken Object Level Authorization has occupied the top position in the
OWASP API Security Top 10~\citep{owasp2019} since its inception in 2019
\citep{owasp2023}. The canonical description is straightforward: an API
fails to verify that the requesting user is authorized to access the
specific object they reference. Despite its prominence, the literature
is almost entirely conceptual. The OWASP guidance offers two
illustrative examples, industry posts provide single-case narratives,
and academic work treats access control models abstractly. The closest
precedents are Barabanov et al.~\citep{barabanov2022}, who systematized
IDOR/BOLA attack patterns from grey literature and OpenAPI
specifications to guide automated detection, and Huang et
al.~\citep{huang2024}, who detected BOLA in database-backed application
source code via static taint analysis. Both works approach BOLA from an
engineering detection perspective i.e., inferring what vulnerabilities
could exist from code or specification structure. Neither characterizes
the empirical distribution of real-world BOLA manifestations from
confirmed disclosures, which is the gap this paper addresses. What is
absent is an empirical account of how BOLA actually manifests in
production: what patterns recur, which identifier types are implicated,
what actions attackers take, and which industries are most exposed.

This paper addresses that gap. Our contributions are:
\begin{itemize}
  \item A refined six-family BOLA taxonomy grounded in real
    disclosures, with operational criteria reducing inter-rater
    ambiguity.
  \item A reproducible classification methodology combining
    LLM-assisted classification with structured human review,
    documented for replication.
  \item Quantitative findings across six dimensions: family, action
    type, authorization direction, industry sector, identifier format,
    and exploit mechanism from 84 confirmed in-scope reports.
  \item A characterization of label noise in HackerOne
    IDOR/Improper Access Control tags, showing substantial divergence
    between platform tagging and confirmed in-scope BOLA disclosures.
  \item An open dataset for further research, with classifier prompt,
    rationale and extraction scripts available at
    \url{https://github.com/hackwither/bola-in-the-wild}.
\end{itemize}

The practical motivation is as follows. API security testing platforms
need vocabulary to generate realistic attack scenarios. Developers
learning to prevent BOLA need concrete pattern guidance, not abstract
principles. And researchers building detection tools need an empirical
baseline against which to calibrate. We attempt to supply all three.

\subsection{IDOR and BOLA: Relationship and Terminology}

IDOR (Insecure Direct Object Reference) and BOLA are related but
distinct concepts that are frequently conflated in practitioner
literature and bug bounty tagging.

IDOR describes the mechanism: an attacker directly references an
internal object using an exposed or predictable identifier, which may be
a URL parameter, a request body field, a GraphQL node ID. It was
introduced by OWASP in 2007 to describe a specific attack
technique~\citep{owasp2007}.

BOLA describes the root cause: the server fails to verify that the
requesting user is authorized to access the specific object they are
referencing. BOLA is the OWASP API Security Top 10 framing introduced
in 2019 to capture this class of authorization failure in API contexts.

BOLA is the broader concept. IDOR is one common technique through which
BOLA is exploited, but BOLA can also be exploited without a direct or
predictable reference, as in chained disclosure attacks where the
identifier is harvested rather than guessed. This paper uses BOLA as
the primary term. Source data was drawn from reports tagged IDOR or
Improper Access Control on HackerOne, since both tags map to the same
underlying failure class in practice. This dataset contributes to
validating that mapping empirically.

\section{Background and Related Work}

\subsection{The OWASP Framing}

The OWASP API Security Top 10 has been updated twice, in 2019
\citep{owasp2019} and 2023~\citep{owasp2023}, with API1 (BOLA) retaining
its top position in both editions. It defines BOLA as occurring when an
API ``does not properly validate that the user performing the request
has the required permissions to access or modify the object,''
distinguishing it from Broken Function Level Authorization (BFLA, \#5)
on the basis that BOLA is object-level; the user has function access but
lacks authorization for the specific resource, while BFLA concerns the
endpoint itself. This distinction breaks down in practice: many real
vulnerabilities involve function access granted but object-level
authorization for state-changing actions absent. Our
``Action-Level Object BOLA'' family captures this hybrid, which
constitutes over a third of the observed cases.

The 2023 revision introduced new attack scenarios, including a GraphQL
deletion example, but the framing remains primarily illustrative rather
than empirically grounded. A 2024 technical report by the Carnegie
Mellon Software Engineering Institute~\citep{cmsei2024} provides a
structured description of BOLA aligned with OWASP, noting that BOLA
exploits ``often happen because of design issues'' rather than purely
implementation failures, which is consistent with the Workflow-Context
and Object Rebinding families identified in this work.

\subsection{Practitioner and Conceptual Taxonomy Work}

Practitioner frameworks have progressively refined the conceptual
landscape of BOLA. PortSwigger's Web Security Academy distinguishes
horizontal from vertical privilege escalation~\citep{portswigger}, a
distinction this paper operationalizes empirically. A more systematic
practitioner effort is Barabanov et al.~\citep{barabanov2022}, who
reviewed bug bounty reports and conference literature to derive
IDOR/BOLA ``attack groups'' defined by endpoint properties and
manipulation technique, then mapped these groups to OpenAPI specification
fields to enable automated detection. Their systematization by endpoint
type, HTTP method, and identifier structure is the closest prior work to
our taxonomy. A key difference is scope and orientation: Barabanov et
al.\ enumerate exploitable structural patterns in API specs rather than
confirmed real-world exploitation patterns, and their grouping is not
tested against a labeled disclosure corpus. Our paper contributes the
missing empirical validation layer.

The CWE taxonomy provides formal classification anchors. CWE-639
(Authorization Bypass Through User-Controlled Key) is IDOR's primary
MITRE mapping; its parent, CWE-284 (Improper Access Control), and child,
CWE-566 (Authorization Bypass Through User-Controlled SQL Primary Key),
locate the vulnerability within a broader weakness hierarchy~\citep{cwe639}.
This paper's six families cut across these CWE boundaries: Action-Level
Object BOLA often maps to CWE-639 but is mechanistically distinct from
Direct Object Reference BOLA; Object Rebinding, which exploits
client-supplied ownership fields, has no direct CWE equivalent,
suggesting the formal taxonomy has gaps that real-world disclosures
expose.

Industry measurement reports also provide context. Imperva's 2024 State
of API Security report~\citep{imperva2024} identifies BOLA as one of its
top risk categories, noting an average of 613 APIs per organization with
potential BOLA exposure, but does not break down disclosure populations
by family, exploit mechanism, or identifier type. Such reports motivate
but do not replace empirical analysis.

\subsection{Bug Bounty Disclosures as an Empirical Source}

The use of HackerOne disclosures as a primary data source positions this
paper within a growing body of research that treats bug bounty corpora
as a research instrument. Walshe and Simpson~\citep{walshe2020} and
Malladi and Subramanian~\citep{malladi2020} study the economics and
operational characteristics of bug bounty programs, finding that
program-operating cost is low relative to other security investments and
that both participation and valid report rates are sensitive to program
design. These findings support the assumption that HackerOne disclosures
represent genuine confirmed vulnerabilities; each report in this corpus
was validated by the program's security team before disclosure,
satisfying the empirical source quality criteria discussed in
Section~\ref{sec:method-datasource}.

The label noise finding of this paper (Section~\ref{sec:labelnoise}) that
a substantial divergence exists between HackerOne IDOR/IAC tagging and
ultimately confirmed in-scope BOLA disclosures, connects to a broader
literature on measurement error in security data. Work on noisy labels
in vulnerability prediction datasets~\citep{croft2022} shows that
mislabeled instances in security corpora have systematic origins
(dormant bugs, wrong-class assignment, vague descriptions) and
meaningfully affect both model training and ground-truth statistics.
This paper's characterization of tag-space label noise in HackerOne is,
to our knowledge, the first such measurement for the IDOR/IAC tag
category specifically, and has direct practical implications for any
researcher or practitioner using HackerOne tag counts to estimate BOLA
exposure.

\section{Methodology}
\label{sec:methodology}

\subsection{Data Source: HackerOne}
\label{sec:method-datasource}

HackerOne is a vulnerability disclosure and bug bounty platform where
organizations invite security researchers to identify and report
vulnerabilities in their products. Researchers submit reports detailing
reproduction steps, impact, and evidence, and program teams triage,
validate, and remediate accepted reports. Once remediated, reports are
publicly disclosed at the program's discretion, creating a corpus of
validated real-world vulnerability findings.

This makes HackerOne disclosures a methodologically sound empirical
source for several reasons. Each included report represents a confirmed
vulnerability that was validated by the program's security team before
disclosure. Reports span a wide range of industries, architectures, and
development maturity levels. The disclosure corpus is large enough to
support sampling at scale, and the structured weakness tagging system
enables reproducible query construction.

The primary limitation of this source is selection bias: only programs
that operate public bug bounty programs and choose to disclose reports
publicly are represented. This over-represents developer-tooling
platforms with mature security programs and under-represents fintech,
healthcare, and enterprise-internal APIs, which we address in
Section~\ref{sec:limitations}.

\subsection{Sampling Frame}

We queried the HackerOne Hacktivity public disclosure dataset for all
reports matching: weakness tag ``IDOR'' or ``Improper Access Control'';
disclosure period January 2021--January 2026; sorted by upvotes
descending; first 200 results. This produced 200 candidates: 138 tagged
Improper Access Control and 62 tagged IDOR, spanning 79 medium, 45 low,
42 high, 24 critical, and 10 none severity ratings.

\subsection{LLM-Assisted Classification: Methodological Basis}

The use of an LLM as a schema-completion assistant for vulnerability
classification follows an emerging pattern in security research.
Marchiori et al.~\citep{marchiori2025} demonstrate that LLMs can learn
vulnerability evaluation from historical assessments with performance
comparable to human experts when guided by ontology-enriched prompts.
The present paper applies this line of work differently: rather than
generating novel classifications, the LLM is used to complete structured
extraction schemas given a fixed taxonomy with operational criteria
(Appendix~\ref{app:prompt}), with all low-confidence outputs reviewed
and adjudicated by human raters. This constrained schema-completion use
is methodologically more conservative than open-ended LLM
classification and is a closer analog to the ``analyst-in-the-loop''
configurations advocated in recent applied work~\citep{farr2025}. The
override and rationale traces provide partial transparency into
classifier reliability, though they do not eliminate the systematic
errors noted in Section~\ref{sec:limitations}.

\subsection{Pre-Filter and Report Acquisition}

All 200 candidates were evaluated against three binary inclusion
criteria using an LLM-assisted pre-filter: (1)~a specific concrete
object is referenced, not generic summarised ``data''; (2)~evidence
exists of cross-boundary access to another user's, tenant's, or role's
object; (3)~at least one technical detail is present (endpoint path,
HTTP method, identifier type, or exploitation steps).

Of 200 candidates, 122 passed (61.0\%). The 78 excluded reports
comprised wrong vulnerability class entirely (${\approx}35$); pure auth
bypass without object reference (${\approx}15$); business logic on own
resources (${\approx}10$); too vague (${\approx}6$); and
misconfiguration/disclosure (${\approx}12$). For the 122 eligible, full
report content was fetched via the HackerOne JSON endpoint
(\texttt{/reports/\{id\}.json}); reports returning only acknowledgment
text or advisory URLs were classified Unclassified.

\subsection{Taxonomy Development}

The taxonomy was developed iteratively through two phases. In Phase~1,
we manually classified 16 reports against a draft taxonomy, which
surfaced four definitional gaps requiring correction: the Direct Object
Reference definition was narrowed to sequential IDs (fixed to be
identifier-type-agnostic); Workflow-Context BOLA was over-broad
(narrowed to object-state-as-enabling-condition); no explicit RBAC/BFLA
exclusion existed (added); and confidence levels lacked operational
definitions (added three-tier criteria keyed to
endpoint/request/cross-user access presence).

\subsection{Classification Procedure}

Classification used an LLM-assisted schema-completion procedure in
which Claude Sonnet (claude-sonnet-4-6) served as a constrained
extraction assistant. The model was provided with a fully-specified
prompt encoding all family definitions, inclusion criteria, confidence
thresholds, and BFLA exclusion rules developed in Phase~1; it did not
make taxonomy design decisions. Its role was restricted to applying
pre-defined criteria to individual report text and populating a fixed
output schema.

Each report was classified against the six-family taxonomy, producing
per-report outputs for all schema fields. Human adjudication was applied
at two documented intervention points with override authority:
(1)~low-confidence pre-filter decisions (${\approx}20$--25 rows,
manually reviewed with override reasons); and (2)~post-classification
review of flagged rows, low-confidence classifications, Unclassified
in-scope reports, and classifier--manual disagreements. Sparse-content
low-confidence cases were enriched with researcher-published writeups
where available before reclassification. Every intervention is
recorded; classifier rationale is preserved for every automated
decision.

To assess classifier reliability, the first author independently coded
a stratified sample of 20 reports without reference to the classifier
output; agreement on family assignment was 85\% (17/20), with all three
disagreements resolved through the documented override procedure and
concentrated in the Direct Object Reference / Action-Level Object
boundary, which is identified in the taxonomy as the primary ambiguity
zone.

All proportions reported in Sections~\ref{sec:labelnoise}--\ref{sec:robustness}
are accompanied by Wilson score 95\% confidence intervals (CIs),
computed as the score interval with continuity correction $z = 1.96$.
Wilson score intervals are used in preference to Wald intervals because
Wald coverage degrades at small samples and at proportions near 0 or 1,
both of which occur in this dataset. For mechanism frequencies
(Section~\ref{sec:mechanisms}), each CI uses the full $n = 84$ as the
denominator; intervals are non-exclusive and do not imply that the
proportions sum to 100\%. All materials, including the classifier
prompt, extraction scripts, and dataset, are available at
\url{https://github.com/hackwither/bola-in-the-wild}.

\subsection{Final Dataset}

From 200 sampled candidates, 107 reports were fully classified. Of
these, 84 (78.5\%, 95\% CI [69.7\%, 85.3\%]) were confirmed in-scope
BOLA; 23 (21.5\%, 95\% CI [14.7\%, 30.3\%]) were Unclassified with
documented reasons. Reports originated from 29 distinct programs across
6 industry sectors. Confidence distribution: High~42 (50.0\%, 95\% CI
[39.5\%, 60.5\%]), Medium~21 (25.0\%, 95\% CI [16.9\%, 35.3\%]),
Low~21 (25.0\%, 95\% CI [16.9\%, 35.3\%]).

\begin{figure}[htbp]
  \centering
  \includegraphics[width=0.35\linewidth]{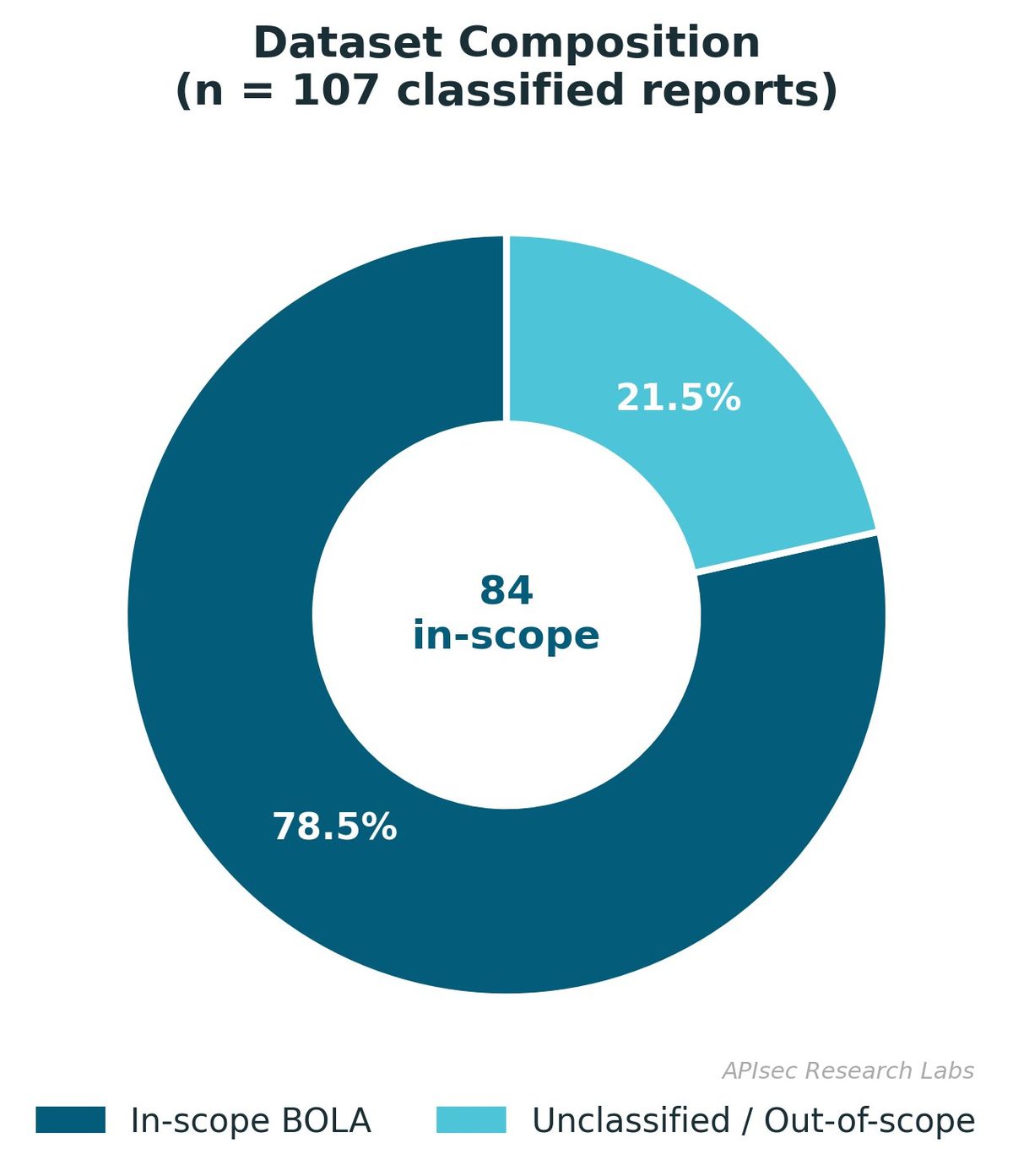}
  \caption{Dataset composition: confirmed in-scope BOLA reports
    (84, 78.5\%) vs.\ unclassified or out-of-scope (23, 21.5\%)
    of 107 fully classified candidates.}
  \label{fig:composition}
\end{figure}

\begin{figure}[htbp]
  \centering
  \includegraphics[width=0.35\linewidth]{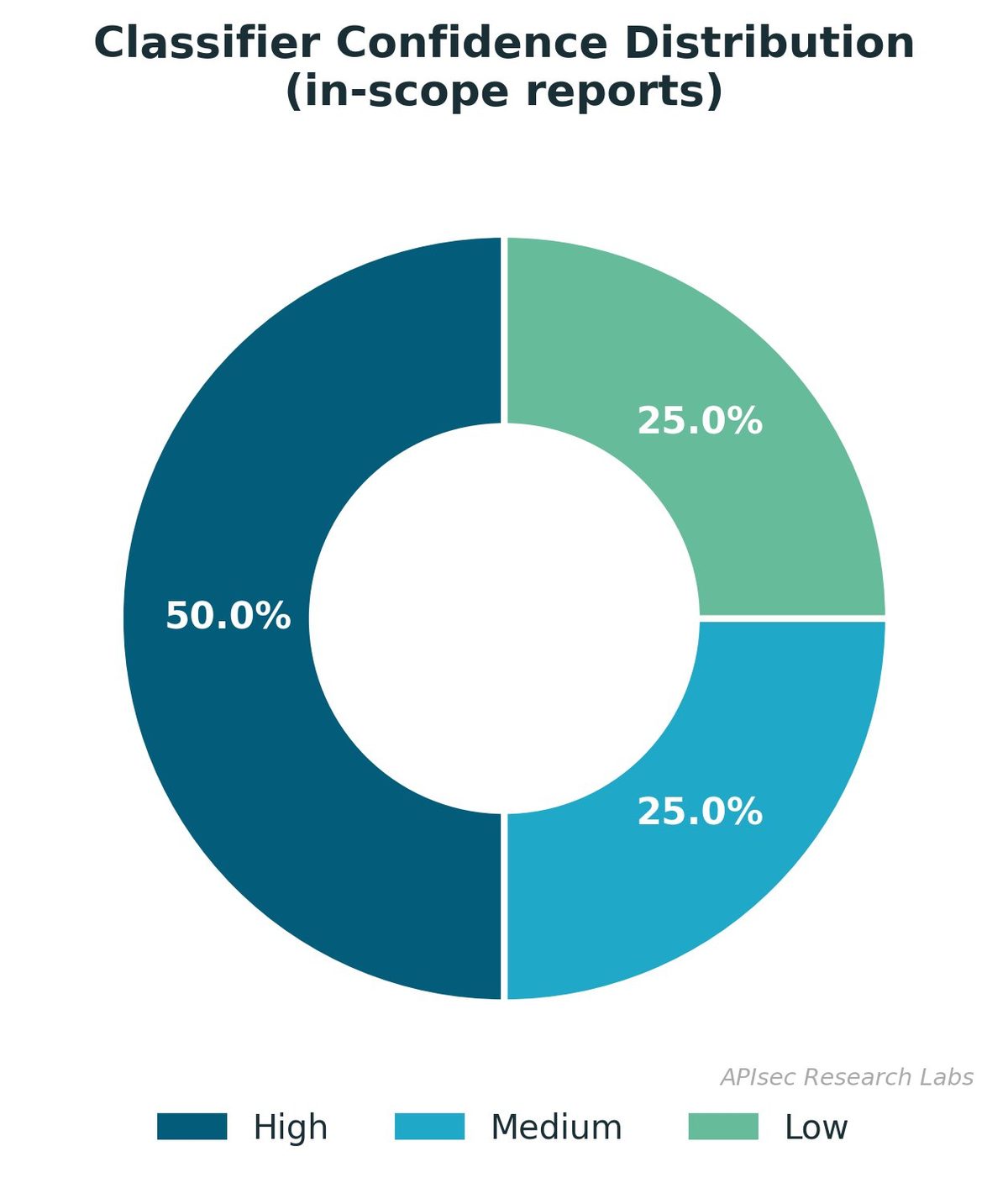}
  \caption{Classifier confidence distribution across 84 confirmed
    in-scope reports (High 50.0\%, Medium 25.0\%, Low 25.0\%).}
  \label{fig:confidence}
\end{figure}

\section{BOLA Taxonomy}
\label{sec:taxonomy}

We define six families, each characterized by its primary authorization
failure mechanism. A report is assigned to the family that best
describes why the authorization failed. Families are defined by the
primary structural or contextual condition enabling the authorization
failure, not by endpoint semantics, HTTP method, or identifier format.
This means that families such as Chained Disclosure, Workflow-Context,
and Object Rebinding describe preconditions under which an authorization
check is absent or stale; they are distinct authorization classes
because the failure only exists given a specific enabling condition that
a correctly-implemented server would enforce independently. Report
references throughout this section use HackerOne report numbers
(e.g., \#3382343). Each is publicly accessible at
\texttt{hackerone.com/reports/\{number\}} for readers wishing to review
primary sources.

\subsection{Direct Object Reference BOLA}

The attacker directly supplies or modifies an object identifier of any
type: sequential integer, UUID, hash, encoded ID, email, or
username, and accesses another user's object without authorization in
a single request. The key criterion is that the identifier was already
known or trivially predictable, requiring no prior acquisition step.
Think of an invoicing app where your invoice lives at /invoices/1041:
you change the number to 1042, and your colleague's invoice loads. The
server confirmed you were logged in but never asked whether 1042 belongs
to you.

\noindent\textit{Example:}
\texttt{GET /ocs/v2.php/apps/dav/api/v1/outOfOffice/\{userId\}}, attacker
substitutes another user's username to retrieve private out-of-office
data including travel destination and phone number (Nextcloud
\#3382343).

\subsection{Action-Level Object BOLA}

The attacker performs an unauthorized state-changing action on another
user's object: delete, modify, approve, transfer, trigger. Read-only
access is Direct Object Reference BOLA regardless of downstream
potential. The key signal is an action verb in the description where
the attacker deletes, modifies, or triggers something belonging to
someone else. The pattern is most visible when you imagine a project
management app where you can archive your own tasks via POST
/tasks/\{id\}/archive: substitute a colleague's task ID and the server
approves the request. You're allowed to call that endpoint, it just
never checked whether the task belongs to you.

\noindent\textit{Example:}
\texttt{POST /v1/account/destroy}, attacker sends victim's email in
the request body using their own session; server does not verify
session-email binding, deleting the victim's account (Mozilla
\#3154983).

\subsection{Tenant Isolation BOLA}

The attacker crosses an organizational/tenant/workspace boundary to
access objects belonging to a different tenant. The violated boundary
is organizational membership specifically, not just user identity. If
User~A reads User~B's data, that is Direct Object Reference; if
Organization~A reads Organization~B's data, that is Tenant Isolation.
A useful way to picture it: a SaaS platform scopes data under a
company\_id, your session belongs to Company A, and you substitute
Company B's identifier into a request body. The server validates that
your session is legitimate but not that it belongs to Company B.

\noindent\textit{Example:}
On HackerOne's own platform, an attacker passed an arbitrary
\texttt{organization\_id} to \texttt{POST /bugs.json} and read private
bug reports from any organization without membership verification
(\#2487889).

\subsection{Workflow-Context BOLA}

Authorization is bypassed specifically because of the object's state or
lifecycle position: draft, archived, deactivated, deleted, transferred.
The same request would be rejected against a normal-state object.
Intentionally narrow: state must be the enabling condition, not a
contributing factor. A CRM is a natural illustration: when a staff
member leaves a company their access is revoked, but the customer
records endpoint only checks whether the requester holds an active staff
session, not whether their access to this specific customer was ever
revoked. The session is still valid; the object-level revocation was
never enforced. This is architecturally distinct from Direct Object
Reference BOLA because a standard ownership check on the same endpoint
would pass under normal object state; the failure is conditional on the
object's lifecycle position, not on identifier predictability.

\noindent\textit{Example:}
Staff accesses a removed customer's personal information after
dissociation from the company, the removal state should revoke access
but does not (Shopify \#2855610).

\subsection{Chained Disclosure BOLA}

Exploitation requires a distinct prior step to harvest the object
identifier from a separate endpoint or workflow step. The attack reads
as a two-act sequence: in the first, the attacker observes a response
that leaks an identifier scoped to someone else's object; in the
second, they present that identifier at a different endpoint that
performs no cross-user ownership check. A healthcare portal provides an
intuitive example: a patient messaging endpoint leaks an internal
attachment token in a response preview, and that token can later be
supplied to a download endpoint to retrieve another patient's medical
document because the second endpoint validates the token but not who it
belongs to. The authorization class is the absent ownership check at
the consuming endpoint, not the cross-endpoint identifier acquisition;
the latter is the precondition that makes the failure exploitable, not
the failure itself.

\noindent\textit{Example:}
In a confirmed case against a U.S.\ Department of Defense system, an
attacker first sent a crafted payload to a Salesforce Aura API endpoint
and received 2,000 ContentDocument IDs belonging to other users'
uploaded files. Each ID, obtained only through that initial bulk query,
was then used individually to download the files via a second endpoint
that validated the document reference but performed no ownership check
against the requesting user's session (\#2623715).

\subsection{Object Rebinding BOLA}

The attacker modifies an ownership-identifying field in the
request, whether \texttt{owner\_id}, \texttt{account\_id},
\texttt{user\_id}, \texttt{msg.Sender} which causes the object to change
ownership or authorization scope. The exploit targets the field that
determines who owns the object, not merely which object to access.
Such a scenario may demonstrate this: a cloud collaboration app lets
users update document metadata via PATCH /documents/\{id\}. The request
body includes an owner\_id field intended for internal use. By replacing
owner\_id with another user's identifier, the attacker reassigns the
document into a different ownership context because the server trusts
the client-supplied ownership field. The authorization failure is the
server's trust in client-supplied ownership metadata; the field
manipulation is the exploit mechanic, but the root cause is the absence
of session-derived ownership verification.

\noindent\textit{Example:}
Attacker packs \texttt{msg.Sender} (victim's address) into a Cosmos SDK
\texttt{MsgExecute}; \texttt{checkSender} validates
\texttt{msg.Sender} instead of authenticated context, triggering fund
transfer from victim's account (Cosmos \#2976481).

\subsection{Unclassified}

Insufficient evidence to confirm the inclusion rule, or explicitly
excluded by the BFLA/auth-bypass/business-logic exclusion criteria.
Each Unclassified report has a documented reason in the classifier
rationale.

\section{Results}
\label{sec:results}

Results are reported across six dimensions: family distribution,
authorization direction, action type, industry sector, identifier
format, and exploit mechanism. Each section opens with a definition of
the dimension being measured. All percentages are calculated over the
\textbf{84 confirmed in-scope reports} unless stated otherwise.

\begin{figure}[htbp]
  \centering
  \includegraphics[width=\linewidth]{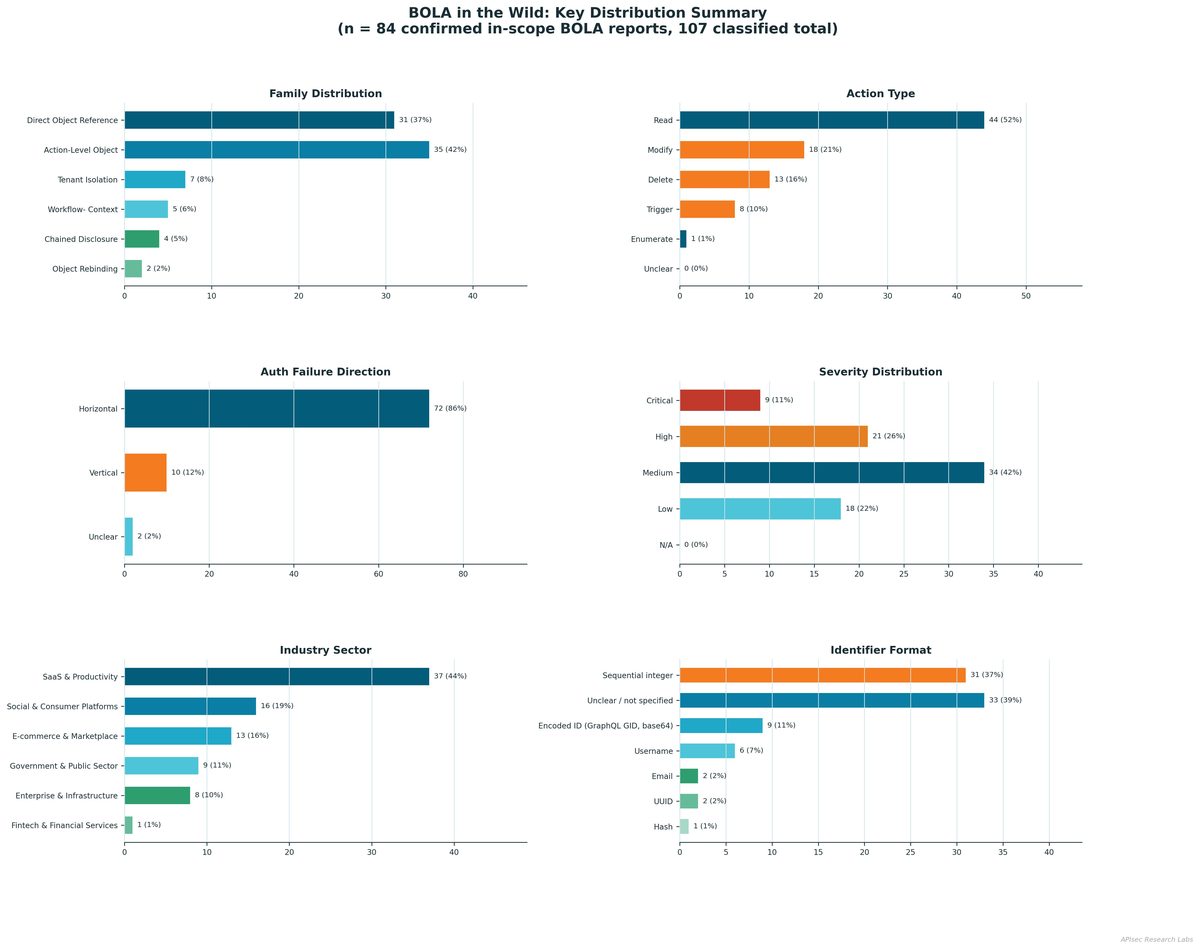}
  \caption{Results overview dashboard across all six analytical
    dimensions ($n = 84$ confirmed in-scope BOLA reports,
    107 classified total).}
  \label{fig:dashboard}
\end{figure}

\subsection{Label Noise in the IDOR/IAC Tag Space}
\label{sec:labelnoise}

Before examining confirmed BOLA patterns, it is worth characterizing
the quality of the source tags themselves. A meaningful share of reports
labeled IDOR or Improper Access Control on HackerOne do not describe
BOLA under rigorous criteria; understanding this noise is itself a
finding with practical implications for how practitioners use tag-based
metrics.

Of the 200 tagged reports sampled, 78 (39.0\%) were excluded during
pre-filtering and a further 23 eligible reports remained unclassified
after full review. Leading exclusion classes: entirely wrong
vulnerability class (XSS, SSRF, CVEs, subdomain takeover:
${\approx}35$, 44.9\%); pure auth bypass without object reference
(${\approx}15$, 19.2\%); business logic on own resources
(${\approx}10$, 12.8\%); and too vague or generic misconfiguration
(${\approx}13$, 16.7\%). Among the 107 fully classified reports, an
additional 23 (21.5\%, 95\% CI [14.7\%, 30.3\%]) were out-of-scope.
Consequently, only 84 of 200 sampled reports (42.0\%) were ultimately
confirmed as in-scope BOLA. The 39.0\% figure should therefore be
interpreted as pre-filter label noise rather than total non-BOLA
prevalence within the sampled tag space. Together, these findings
suggest that a meaningful proportion of practitioner-labeled `IDOR' is
not BOLA by rigorous definition, so organizations using tag-counting
to assess access-control exposure will significantly overstate the
BOLA-specific signal.

\subsection{BOLA Family Distribution}
\label{sec:families}

This section reports how confirmed BOLA cases distribute across the six
taxonomy families. The distribution characterizes which authorization
failure mechanisms are most prevalent in real-world disclosures.
Among the 84 in-scope reports (see Figure~\ref{fig:families}):

\begin{figure}[htbp]
  \centering
  \includegraphics[width=0.75\linewidth]{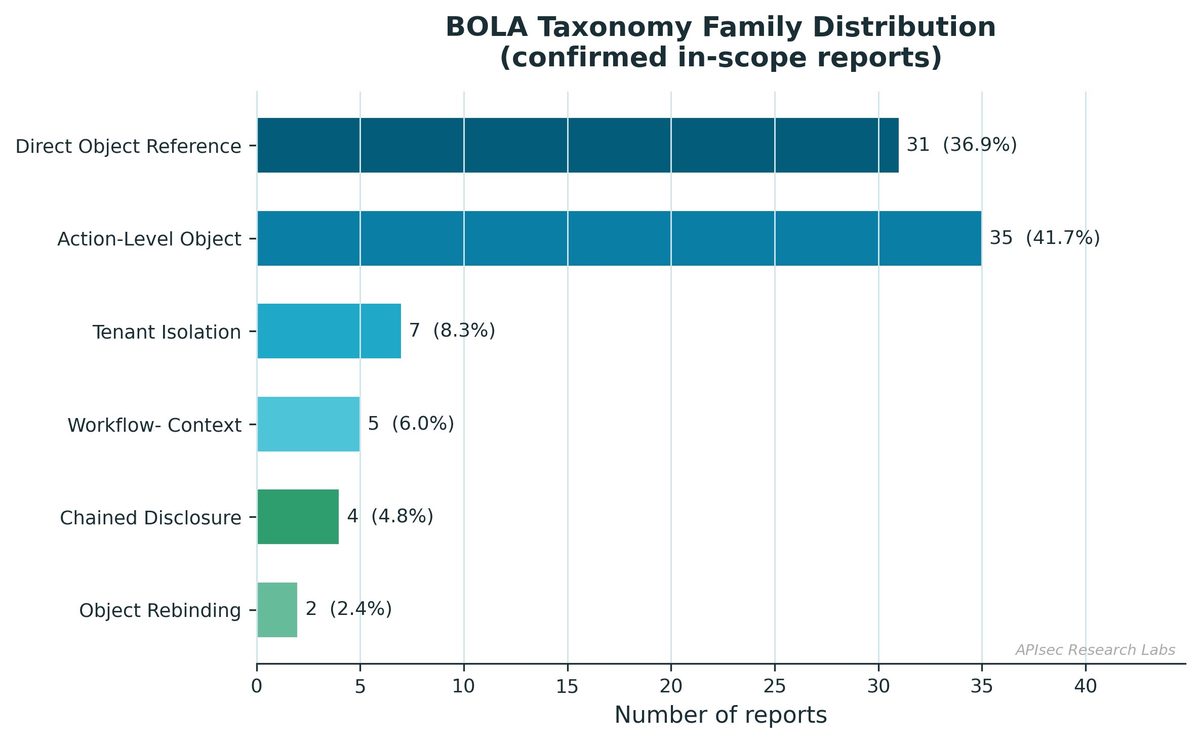}
  \caption{BOLA taxonomy family distribution across 84 confirmed
    in-scope reports. Action-Level Object BOLA (41.7\%) and Direct
    Object Reference BOLA (36.9\%) are the two dominant observed
    families.}
  \label{fig:families}
\end{figure}

\begin{table}[htbp]
  \centering
  \caption{BOLA Family Distribution ($n = 84$)}
  \label{tab:families}
  \begin{tabular}{lrrl}
    \toprule
    \textbf{BOLA Family} & \textbf{Count} & \textbf{\%} & \textbf{95\% CI} \\
    \midrule
    Action-Level Object BOLA     & 35 & 41.7\% & {[}31.7\%, 52.4\%{]} \\
    Direct Object Reference BOLA & 31 & 36.9\% & {[}27.4\%, 47.6\%{]} \\
    Tenant Isolation BOLA        &  7 &  8.3\% & {[}3.8\%, 16.5\%{]}  \\
    Workflow-Context BOLA        &  5 &  6.0\% & {[}2.2\%, 13.5\%{]}  \\
    Chained Disclosure BOLA      &  4 &  4.8\% & {[}1.5\%, 12.0\%{]}  \\
    Object Rebinding BOLA        &  2 &  2.4\% & {[}0.1\%, 8.8\%{]}   \\
    \midrule
    Total                        & 84 & 100\%  & ---                  \\
    \bottomrule
  \end{tabular}
\end{table}

Action-Level Object BOLA leads the distribution at 41.7\% (95\% CI
[31.7\%, 52.4\%]), although its lead over Direct Object Reference BOLA
is sensitive to program weighting and not statistically distinguishable
given the overlap in confidence intervals. This represents cases
where the attacker acts on another user's data like deleting, modifying,
triggering workflows, rather than merely reading it. A
program-weighted robustness check (Section~\ref{sec:robustness}) shows
this lead is sensitive to HackerOne concentration; under equal program
weighting the two families are effectively co-dominant (Action-Level
32.4\%, Direct Object Reference 37.5\%). The confidence intervals of
the two leading families overlap substantially: Action-Level
[31.7\%, 52.4\%] versus Direct Object Reference [27.4\%, 47.6\%],
providing statistical grounding for the co-dominance interpretation
independent of the program-weighting analysis.

Direct Object Reference BOLA follows at 36.9\%, confirming that direct
identifier substitution remains a prevalent real-world BOLA vector
despite decades of developer education, and not only in unsophisticated
programs: it appears across HackerOne's own platform, GitHub, U.S.\
Department of Defense (DoD) systems, and Shopify. Together,
Action-Level and Direct Object Reference account for 78.6\% of the
observed population. The remaining four families collectively represent
21.4\% and tend toward greater exploit complexity, which may explain
lower prevalence or may reflect underreporting due to harder
reproduction.

\subsection{Authorization Failure Direction: Horizontal vs.\ Vertical}
\label{sec:direction}

Authorization failures can run in two directions: \emph{horizontal},
where a user accesses an object belonging to a peer at the same
privilege level, and \emph{vertical}, where a lower-privileged user
accesses an object owned by a higher-privileged role such as an
administrator. Vertical is not a role-gating failure; in every vertical
case in this dataset, the attacker's role was legitimately permitted to
call the endpoint. The missing control was ownership of the specific
object, not access to the function. Vertical BOLA is not the same as
RBAC failure or BFLA, where the issue is role-gated endpoint access
rather than object ownership.

\begin{figure}[htbp]
  \centering
  \includegraphics[width=0.7\linewidth]{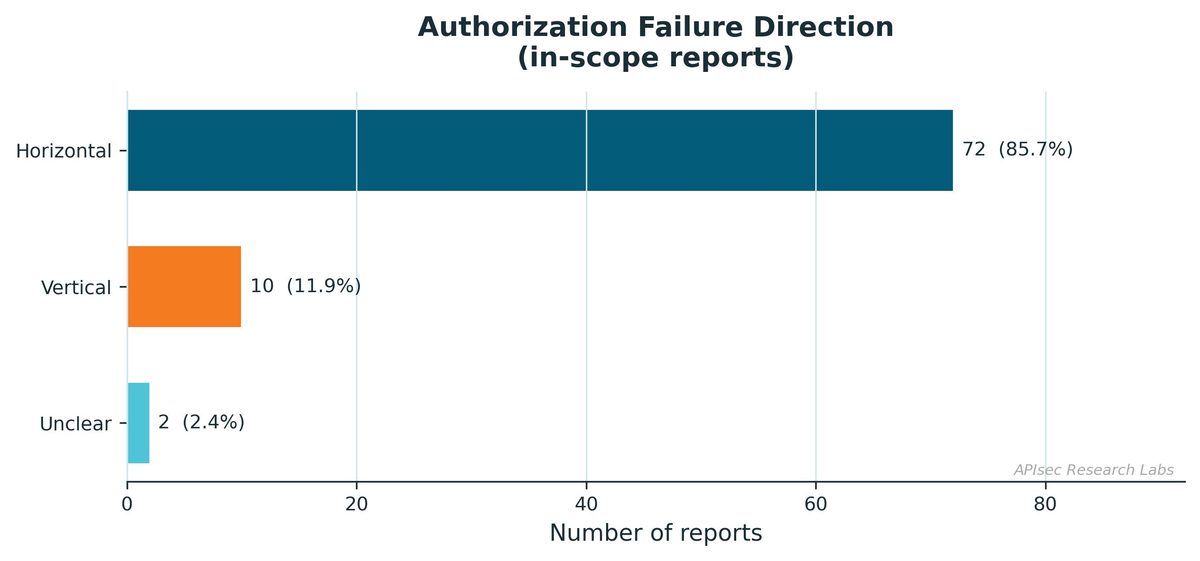}
  \caption{Authorization failure direction across 84 confirmed
    in-scope reports. Horizontal failures dominate at 85.7\%
    (95\% CI [76.5\%, 91.8\%]) while vertical failures account
    for 11.9\% (95\% CI [6.4\%, 20.7\%]).}
  \label{fig:direction}
\end{figure}

Horizontal failures dominate at 85.7\% while vertical failures account
for 11.9\% and are disproportionately underrepresented in developer
guidance, which typically frames BOLA as peer-to-peer. Vertical cases
include a standard Nextcloud user deleting an admin's external storage
(report \#2212627), a Frontegg admin modifying an Owner's API key via
an undocumented PATCH method (\#2149124), and multiple Autodesk users
gaining modification capability over admin-owned objects. Vertical
failures are particularly high-impact because they allow a standard
user to access or modify objects owned by administrator accounts,
gaining effective administrative reach without being granted an
administrative role.

\subsection{Action Type Distribution}
\label{sec:actiontype}

Beyond who owns the object, it matters what the attacker can do to it.
Action type describes the primary unauthorized operation the attacker
performs on the target object.

\begin{itemize}
  \item \textbf{Read:} the attacker retrieves data they are not
    authorized to view.
  \item \textbf{Modify:} the attacker changes fields or properties of
    another user's object.
  \item \textbf{Delete:} the attacker removes another user's object
    entirely.
  \item \textbf{Trigger:} the attacker initiates a state change or
    side effect on another user's object without directly modifying
    its data (e.g., forcing a ride acceptance, sending a message to a
    restricted inbox, or initiating a payment).
  \item \textbf{Enumerate:} the attacker iterates over objects to
    confirm their existence or retrieve metadata without necessarily
    reading full content.
\end{itemize}

\begin{figure}[htbp]
  \centering
  \includegraphics[width=0.75\linewidth]{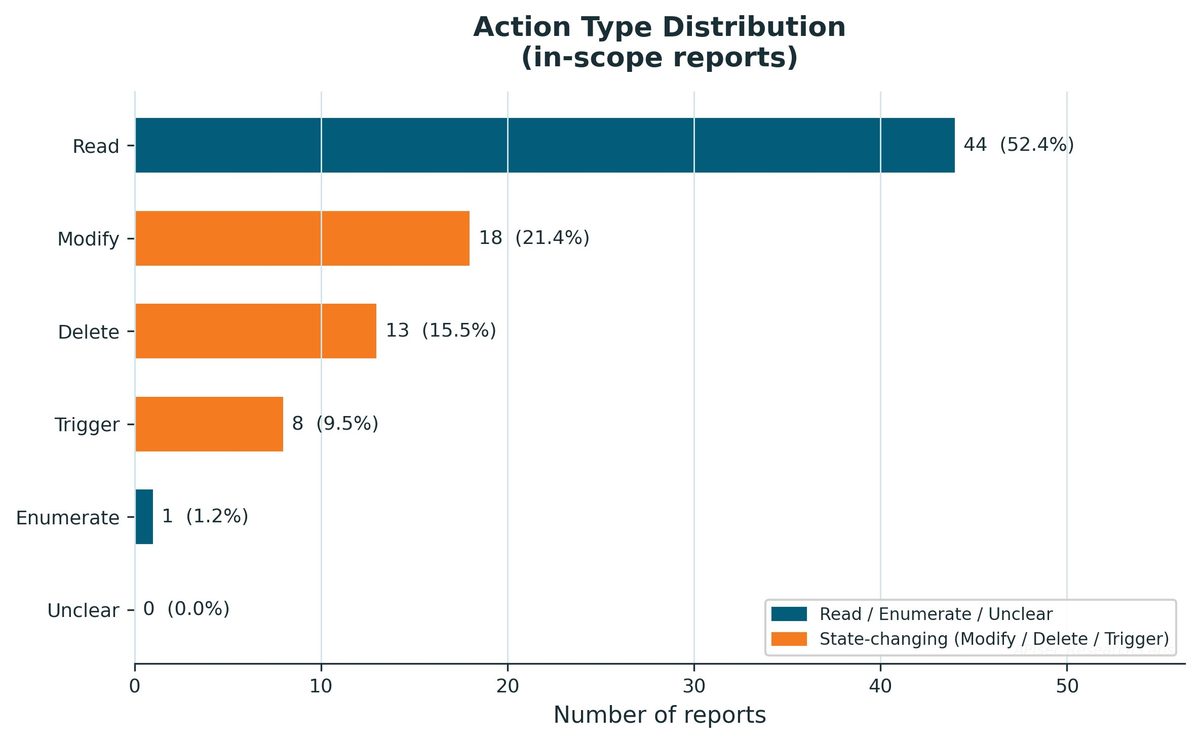}
  \caption{Action type distribution across 84 confirmed in-scope
    reports. State-changing actions (Modify + Delete + Trigger)
    collectively account for 46.4\% of cases.}
  \label{fig:actiontype}
\end{figure}

Unauthorized read (52.4\%) is the most common single action, but
state-changing types combined: Modify + Delete + Trigger = 39 cases
(46.4\%) account for nearly half of all confirmed BOLA. The Delete
type (15.5\%) is particularly notable: deletion attacks are
high-impact, often irreversible, and unlikely to trigger
data-exfiltration monitoring. Reports include deletion of API tokens
(Mozilla), analytics reports (HackerOne), external storage
configurations (Nextcloud), and full user accounts (Mozilla). The
absence of any ``Unclear'' action classifications in this dataset reflects improved report
coverage and more complete enrichment of low-confidence cases.

\subsection{Industry Sector Distribution}
\label{sec:sectors}

Industry sector reflects the primary product category of the
disclosing program, not the parent company. Distribution here is
shaped by both actual vulnerability density and disclosure norms;
sectors with conservative disclosure practices will be
underrepresented relative to their true exposure.

\begin{figure}[htbp]
  \centering
  \includegraphics[width=0.8\linewidth]{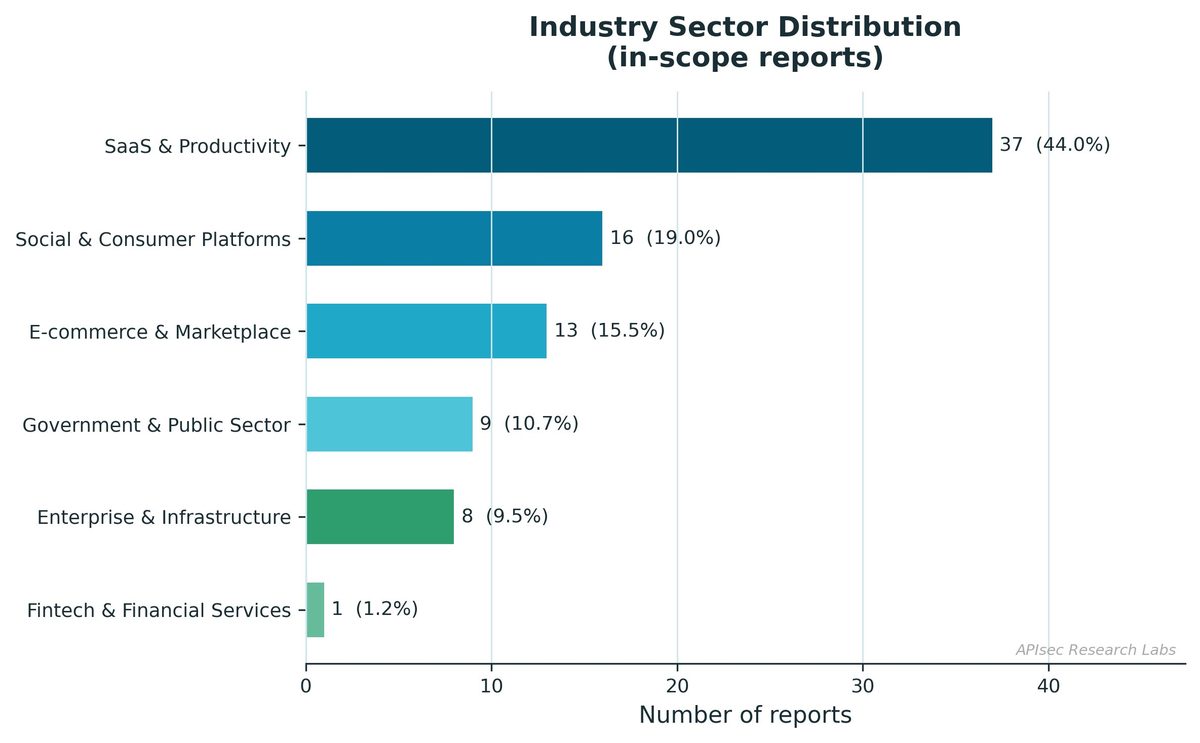}
  \caption{Industry sector distribution across 84 confirmed in-scope
    reports.}
  \label{fig:sectors}
\end{figure}

SaaS \& Productivity dominates (44.0\%), consistent with
developer-facing platforms (GitHub, GitLab, HackerOne, Nextcloud)
operating active public bounty programs. Government \& Public Sector
(10.7\%) produces a disproportionate share of Critical-severity
disclosures: nine DoD reports include exfiltration of Air Force
candidate PII with medical and criminal records (\#2968391, Critical),
SAAR form access with military PII (\#2967032, High), and Salesforce
object abuse via the Aura API (\#2950536, Critical).
Program-weighted reanalysis reduces the Government \& Public Sector
share to 3.6\% (Section~\ref{sec:robustness}), confirming that this
sector's dataset representation is driven by a single concentrated
program; the individual report findings remain valid but do not
generalize to broad sector exposure. Fintech \& Financial Services is
notably underrepresented at 1.2\%, with a single confirmed in-scope report
(Cosmos \#2976481). This is almost certainly a reflection of
conservative disclosure norms and regulatory constraints rather than
lower actual vulnerability density.

\subsection{Identifier Format Distribution}
\label{sec:identifiers}

Identifier format describes the type of object reference used during
exploitation. Sequential integers are incrementable numeric
identifiers; encoded IDs are opaque representations such as
base64-encoded GraphQL Global IDs (GIDs); usernames and emails use
user-derived identity fields as object references; UUIDs are randomly
generated globally unique identifiers; and hashes are deterministic
encoded object references. This distribution helps evaluate whether
non-sequential or opaque identifier schemes meaningfully reduce
real-world BOLA exposure.

\begin{figure}[htbp]
  \centering
  \includegraphics[width=0.75\linewidth]{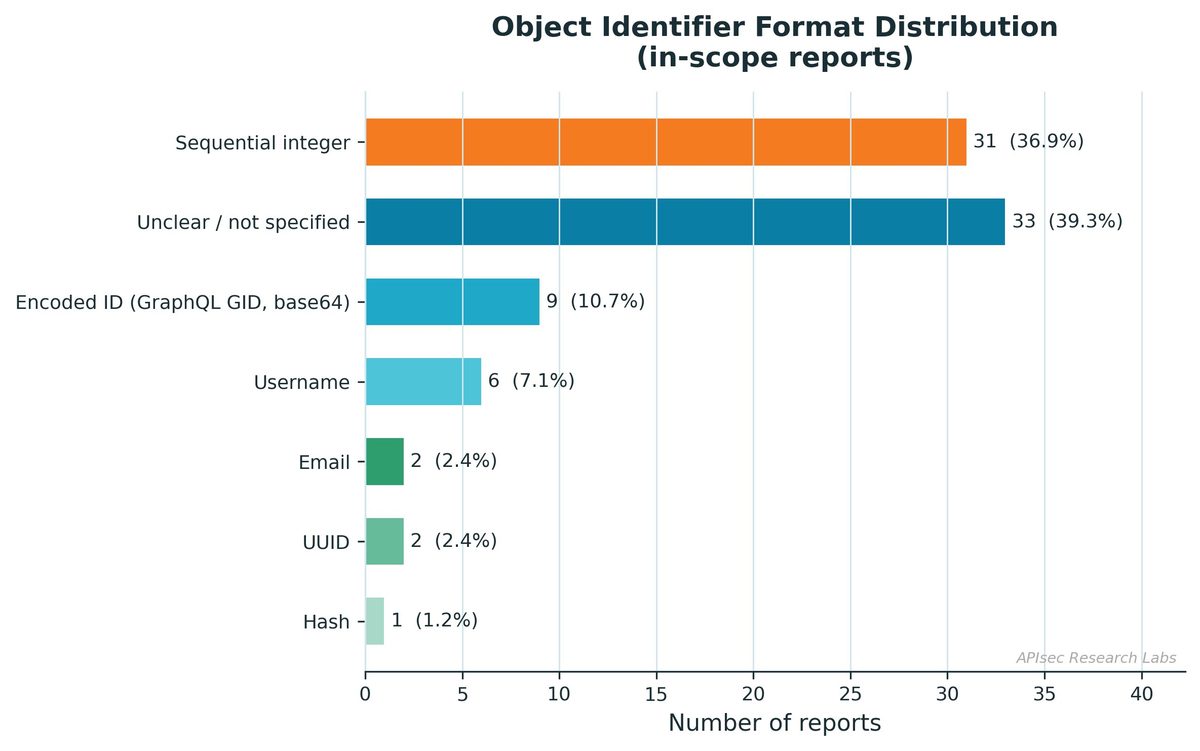}
  \caption{Object identifier format distribution across 84 confirmed
    in-scope reports. Sequential integers lead among known-format
    cases; non-sequential formats account for 39.2\% of known-format
    reports.}
  \label{fig:identifiers}
\end{figure}

Sequential integers remain the most frequently evidenced identifier
type among known-format cases (36.9\% of all in-scope reports),
confirming that predictable IDs represent the highest-risk design
choice. However, non-sequential formats like encoded ID, username, email,
UUID, hash account for 39.2\% of known-format cases, demonstrating
that UUIDs and opaque identifiers do not eliminate BOLA risk.

\paragraph{GraphQL Global IDs (GIDs).}
GIDs are base64-encoded node identifiers used by GraphQL
implementations to create globally unique object references. A typical
GID encodes the service name, object type, and a numeric backend ID.
For example, \texttt{gid://hackerone/Report/3604288} base64-encodes to
the opaque string seen in API responses. Because the underlying numeric
ID is sequential, an attacker who decodes one GID can increment the
integer, re-encode, and reference a different user's object to bypass
the apparent opacity of the identifier.

Encoded IDs (GraphQL Global IDs, base64-encoded node IDs) are
structurally non-guessable but are routinely exploited by decoding the
GID, incrementing the underlying sequential integer, and re-encoding,
as demonstrated against HackerOne (\#2633771), GitLab (\#2528293), and
DoD Salesforce infrastructure (\#2968391). The 39.3\% ``unclear'' rate
reflects sparse HackerOne disclosures from conservative programs, not
an underlying property of the vulnerabilities, and represents an
improvement target for future sampling.

\subsection{Exploit Mechanism Frequency}
\label{sec:mechanisms}

Exploit mechanisms describe the specific technical method used to cross
the authorization boundary. A single report can involve multiple
mechanisms, for example, GraphQL global ID leakage combined with
sequential integer enumeration after decoding. Counts therefore do not
sum to 84.

\begin{figure}[htbp]
  \centering
  \includegraphics[width=0.8\linewidth]{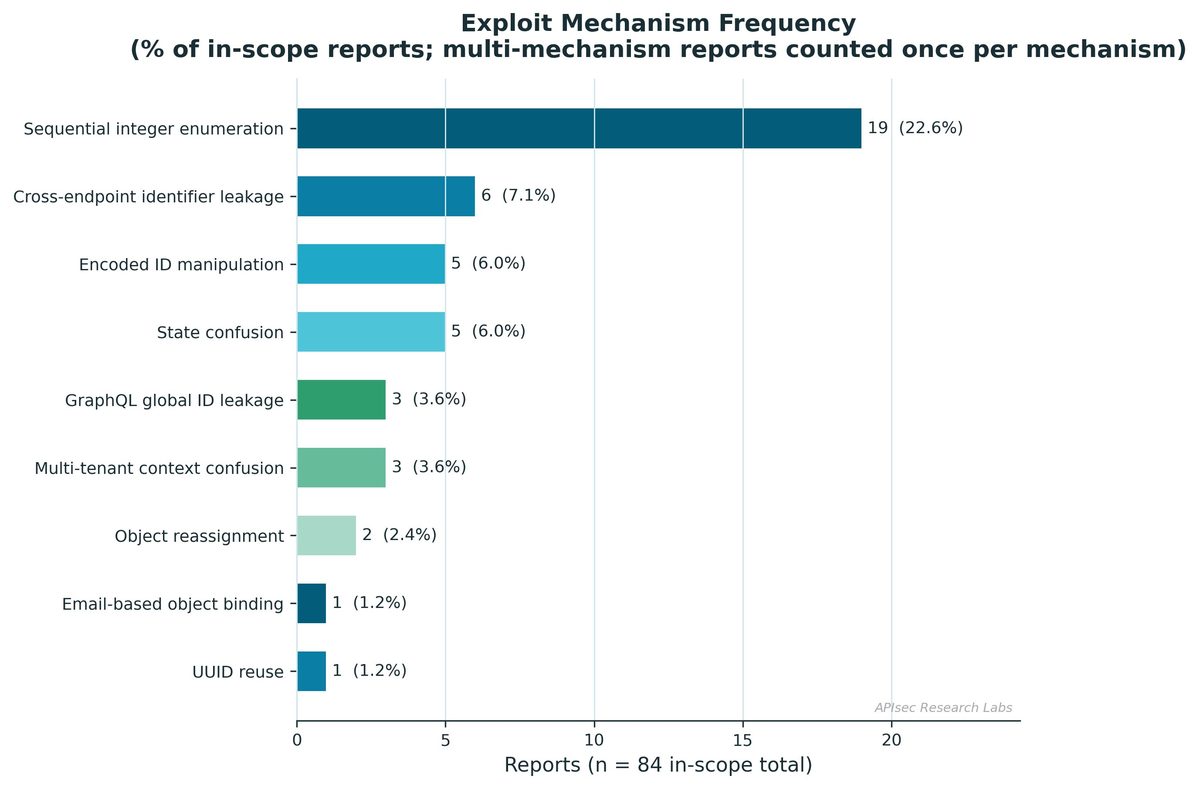}
  \caption{Exploit mechanism frequency across 84 confirmed in-scope
    reports. Multi-mechanism reports are counted once per mechanism;
    counts do not sum to 84.}
  \label{fig:mechanisms}
\end{figure}

\begin{table}[htbp]
  \centering
  \caption{Exploit Mechanism Frequency ($n = 84$ in-scope reports)}
  \label{tab:mechanisms}
  \begin{tabular}{lrrl}
    \toprule
    \textbf{Mechanism} & \textbf{N} & \textbf{\%} & \textbf{95\% CI} \\
    \midrule
    Sequential integer enumeration   & 19 & 22.6\% & {[}14.9\%, 32.7\%{]} \\
    Cross-endpoint identifier leakage &  6 &  7.1\% & {[}3.0\%, 15.0\%{]}  \\
    Encoded ID manipulation           &  5 &  6.0\% & {[}2.2\%, 13.5\%{]}  \\
    State confusion                   &  5 &  6.0\% & {[}2.2\%, 13.5\%{]}  \\
    GraphQL global ID leakage         &  3 &  3.6\% & {[}0.8\%, 10.4\%{]}  \\
    Multi-tenant context confusion    &  3 &  3.6\% & {[}0.8\%, 10.4\%{]}  \\
    Object reassignment               &  2 &  2.4\% & {[}0.1\%, 8.8\%{]}   \\
    UUID reuse                        &  1 &  1.2\% & {[}0.0\%, 7.1\%{]}   \\
    Email-based object binding        &  1 &  1.2\% & {[}0.0\%, 7.1\%{]}   \\
    \bottomrule
  \end{tabular}
\end{table}

Sequential integer enumeration is the single most prevalent mechanism
at 22.6\% (95\% CI [14.9\%, 32.7\%]), persisting in 2023--2026
disclosures from well-resourced organizations. Cross-endpoint
identifier leakage (7.1\%) and the pair of state confusion and encoded
ID manipulation (6.0\% each) together represent patterns that require
multi-step reasoning about identifier provenance and object lifecycle.
This places them architecturally beyond stateless HTTP-level scanners
and pointing toward the need for session-aware, stateful testing.
GraphQL-specific mechanisms (global ID leakage + encoded ID
manipulation) appear in a combined 9.6\% of cases, consistently
exploiting the decode/increment/re-encode pattern across HackerOne,
GitLab, and Shopify.

\subsection{Severity Distribution}
\label{sec:severity}

Severity ratings reflect the disclosing program's assessment at time of
triage, using HackerOne's standard scale: Critical, High, Medium, Low.
These ratings combine impact and exploitability and were not adjusted
in this analysis.

\begin{figure}[htbp]
  \centering
  \includegraphics[width=0.75\linewidth]{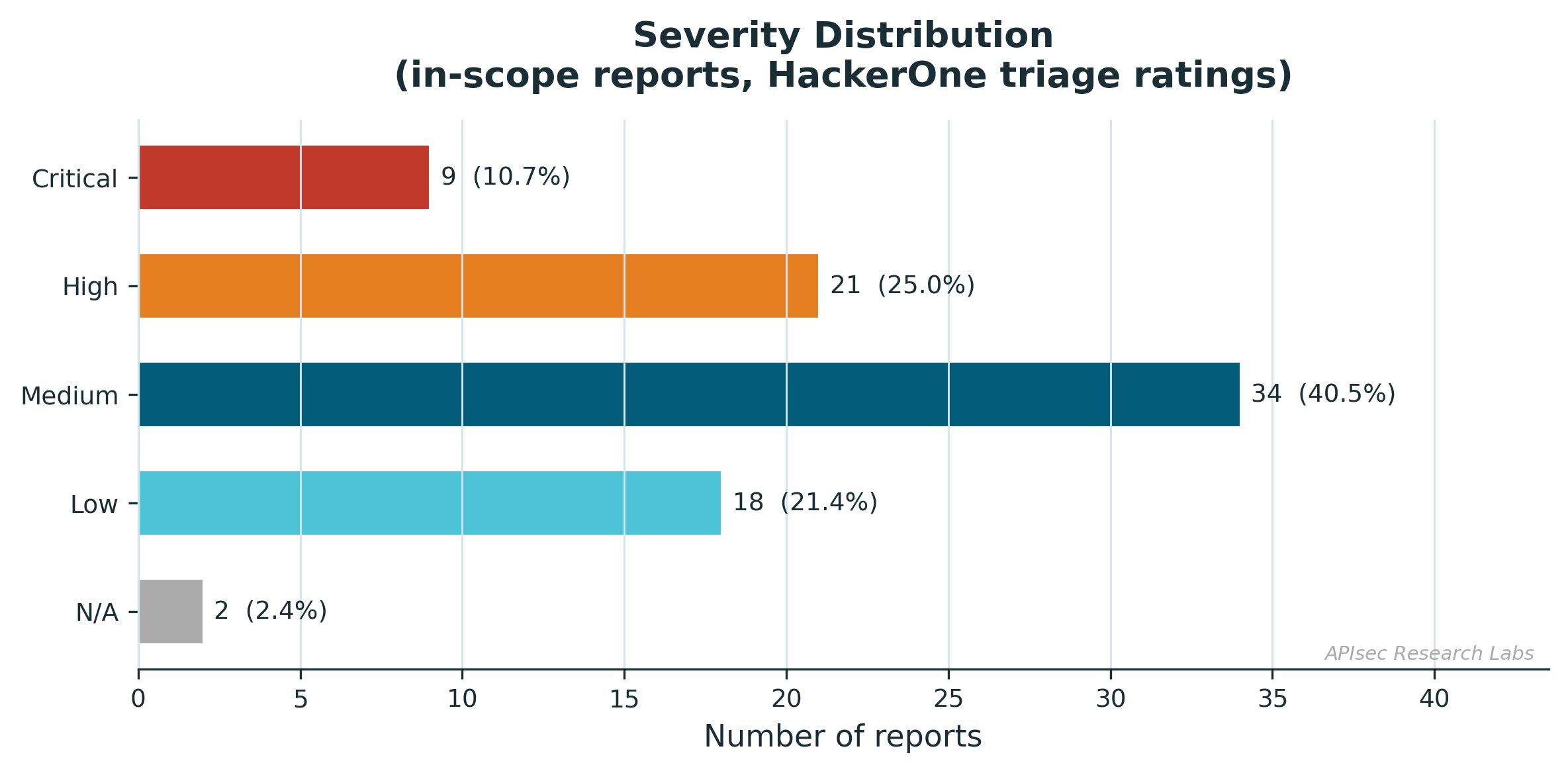}
  \caption{Severity distribution across 84 confirmed in-scope reports.
    35.7\% of reports are rated High or Critical.}
  \label{fig:severity}
\end{figure}

\begin{figure}[htbp]
  \centering
  \includegraphics[width=0.75\linewidth]{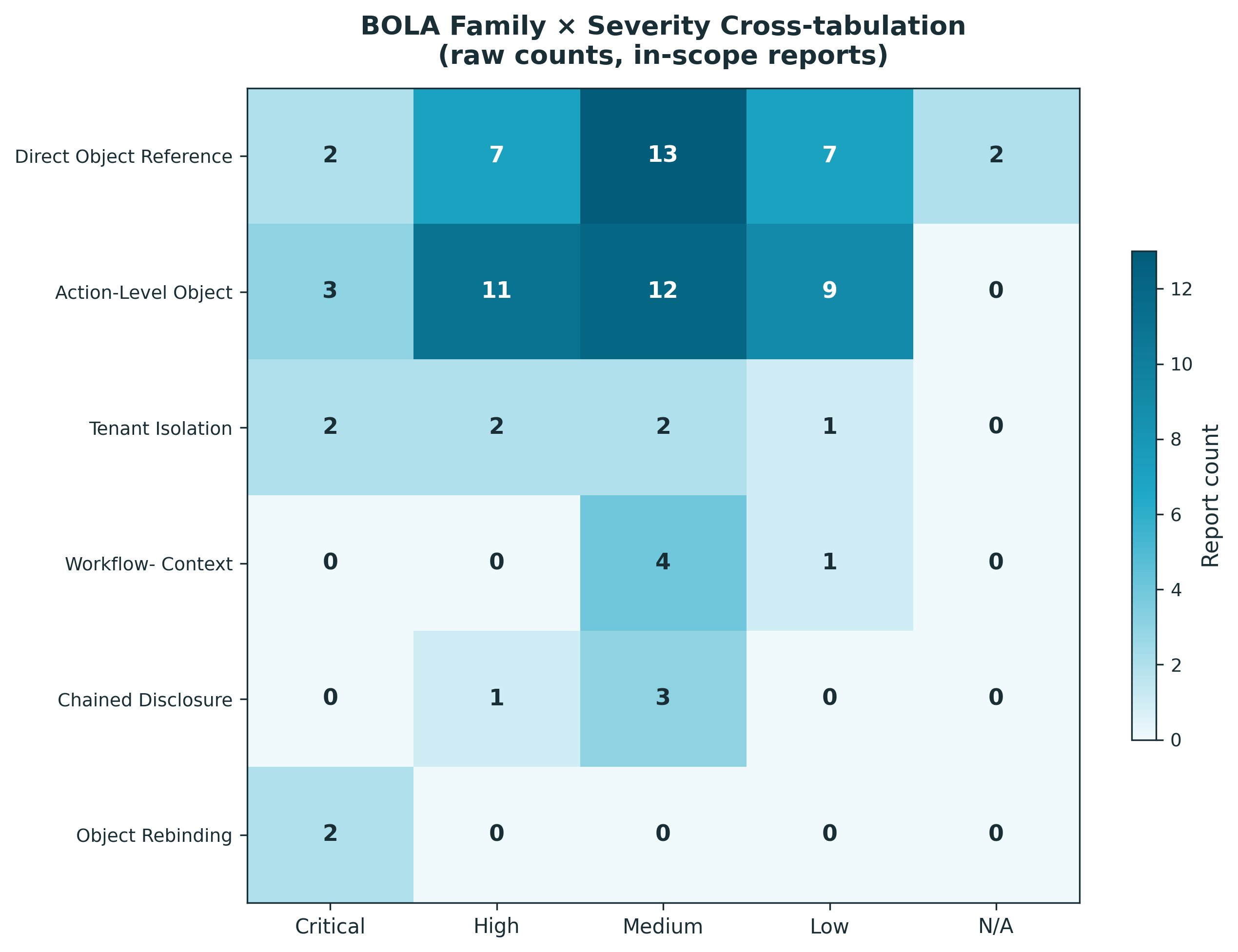}
  \caption{BOLA family $\times$ severity cross-tabulation (raw counts,
    $n = 84$). Action-Level Object BOLA carries disproportionately
    high severity ratings relative to its dataset share.}
  \label{fig:familyseverity}
\end{figure}

The median BOLA report is Medium severity, but 35.7\% (95\% CI
[26.3\%, 46.4\%]) are rated High or Critical. Critical-severity cases
(9 reports, 10.7\%, 95\% CI [5.5\%, 19.3\%]) span the Cosmos SDK
fund-transfer vulnerability (\#2976481), multiple Air Force and DoD
Salesforce data exposures (\#2968391, \#2950536), platform-wide
organizational access on HackerOne (\#2487889), and account
modification on Mars e-commerce (\#2828693, \#2828641). Action-Level
Object BOLA carries disproportionately high severity ratings relative
to its dataset share, reflecting the greater impact of irreversible
state-changing actions compared to read-only disclosure.

\subsection{Temporal Distribution}

Temporal distribution tracks disclosure year across confirmed in-scope
reports. Of 84 in-scope reports, 48 have confirmed disclosure years;
36 (42.9\%) have undisclosed years due to HackerOne redaction, limiting
trend analysis to the dated subset.

\begin{figure}[htbp]
  \centering
  \includegraphics[width=0.78\linewidth]{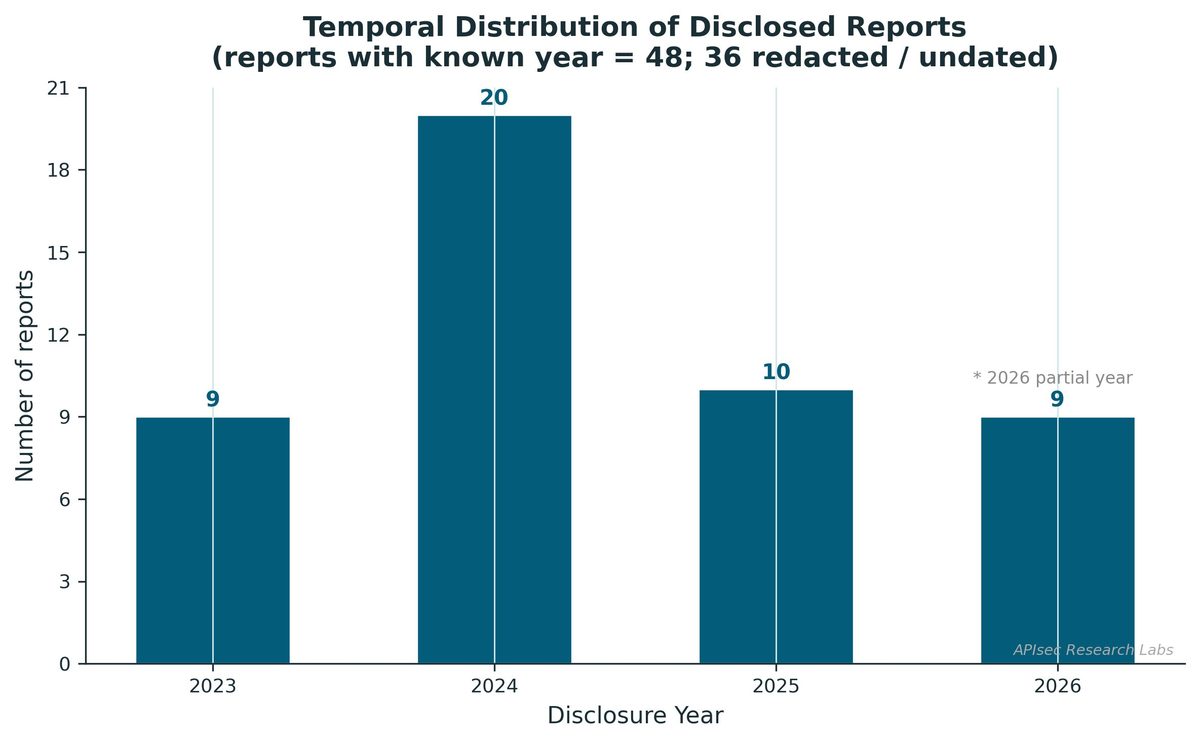}
  \caption{Temporal distribution of confirmed in-scope reports by
    disclosure year. 36 of 84 reports (42.9\%) have redacted or
    undisclosed years and are excluded.}
  \label{fig:temporal}
\end{figure}

The disclosed subset shows no meaningful decline across the 2023--2026
window, which spans the OWASP API Security Top 10 2023 update. The
2024 cohort is the largest dated subset (20 reports), consistent with
increased HackerOne disclosure activity that year. This apparent
stability in vulnerability prevalence suggests the gap between
awareness and remediation remains substantial.

\subsection{Program-Weighted Robustness Analysis}
\label{sec:robustness}

To assess sensitivity to single-program concentration
(Section~\ref{sec:limitations}), each report was reweighted by
$1/n_{\text{program}}$, assigning every program equal total weight of
1.0 regardless of report volume.

\begin{figure}[htbp]
  \centering
  \includegraphics[width=0.75\linewidth]{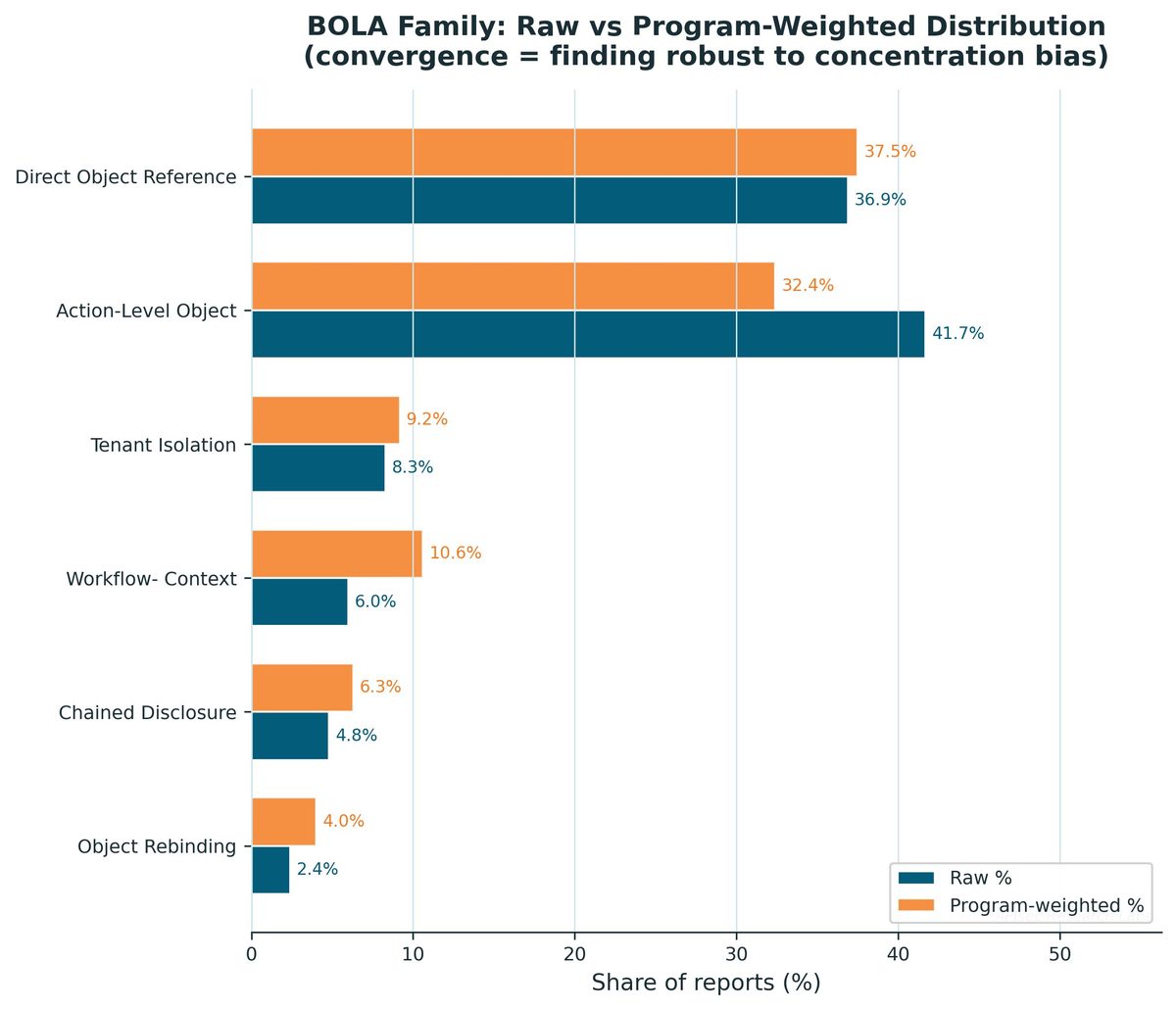}
  \caption{BOLA family distribution: raw count vs.\ program-weighted
    percentage. Convergence indicates a finding robust to
    single-program concentration.}
  \label{fig:weighted_family}
\end{figure}

\begin{figure}[htbp]
  \centering
  \includegraphics[width=0.75\linewidth]{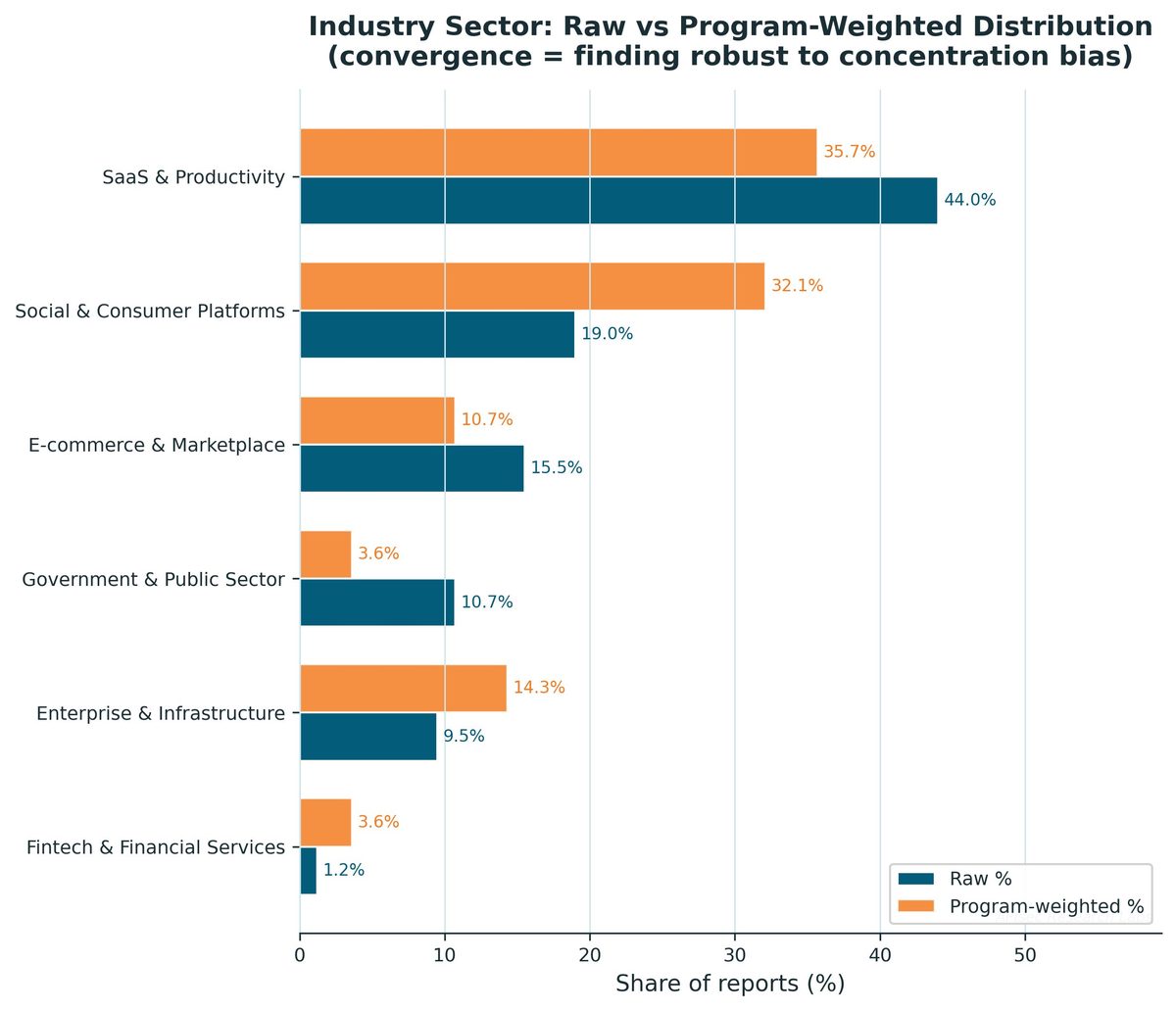}
  \caption{Industry sector distribution: raw count vs.\
    program-weighted percentage. Divergence indicates sectors where
    report volume is concentrated in few programs.}
  \label{fig:weighted_sector}
\end{figure}

\textbf{Family distribution:} Direct Object Reference is stable under
weighting (36.9\% raw, 37.5\% weighted, $+$0.6pp), confirming that
finding is robust to program mix. Action-Level Object contracts by 9.3
percentage points (41.7\% to 32.4\%), attributable primarily to
HackerOne's 13 reports disproportionately representing action-type
violations. Under program-equal weighting the ranking inverts: Direct
Object Reference leads at 37.5\% and Action-Level Object follows at
32.4\%. The two families jointly account for 69.9\% of
program-weighted share versus 78.6\% raw, so their combined dominance
is robust even as the ordering is not. Workflow-Context BOLA shows the
largest proportional increase (6.0\% to 10.6\%, $+$4.6pp), indicating
its raw count understates cross-program prevalence. All other families
shift by less than 2pp.

\textbf{Sector distribution:} Government \& Public Sector is the most
concentration-sensitive cell, contracting from 10.7\% to 3.6\%
($-$7.1pp), directly reflecting DoD's nine reports from a single
program. Social \& Consumer Platforms is the most underweighted sector
in raw counts, rising from 19.0\% to 32.1\% ($+$13.1pp) as
single-report programs in this sector carry equal program weight. SaaS
\& Productivity remains the dominant sector under both methods (44.0\%
raw, 35.7\% weighted), confirming sector-level stability for the
largest category. Enterprise \& Infrastructure rises from 9.5\% to
14.3\% ($+$4.8pp).

These results qualify two specific claims in this paper. First,
Action-Level Object BOLA's lead over Direct Object Reference is
sensitive to HackerOne concentration and should be understood as
co-dominance rather than unambiguous precedence. This interpretation is
reinforced by the Wilson confidence intervals for the raw counts:
Action-Level Object BOLA [31.7\%, 52.4\%] and Direct Object Reference
BOLA [27.4\%, 47.6\%] have overlapping intervals that span roughly 20
percentage points each, consistent with estimation uncertainty at
$n = 84$. The ordering of the two families is not statistically
distinguishable from the raw counts alone.

\section{Discussion}
\label{sec:discussion}

\subsection{Implications for Detection and Testing}

The prevalence of sequential integer enumeration (22.6\%) means an
automated DAST tool with increment/decrement ID fuzzing would detect
roughly a fifth to a quarter of bugs in this population. However,
39.3\% of in-scope reports have undetermined identifier formats, and
non-sequential formats account for 39.2\% of known-format cases.
Effective detection requires semantic understanding of which parameters
are object references, whether server responses differ across users, and
whether the differential constitutes a boundary violation, not just
parameter fuzzing.

The Action-Level Object family (41.7\%) has the most direct
implications for testing protocol. Standard BOLA testing verifies that
Account~B can read Account~A's data. This now misses one of the largest
confirmed families; cases where Account~B can \emph{act on}
Account~A's data. Effective test suites must include delete, modify,
and trigger operations across ownership boundaries as first-class test
cases, not edge cases. A testing framework that only validates
\texttt{GET}-method cross-user access is likely to miss a substantial
part of real-world BOLA patterns.

\subsection{Implications for API Design}

The persistent prevalence of sequential integer IDs (36.9\% of
in-scope reports, 95\% CI [27.4\%, 47.6\%]; 60.8\%, 95\% CI
[47.1\%, 73.0\%] of the 51 reports with a determined identifier
format) from organizations as sophisticated as HackerOne, GitHub,
Mozilla, and U.S.\ Department of Defense agencies in 2023--2026
disclosures demonstrates that sequential integers as
externally-referenced object identifiers remain a live risk in mature
security programs. They make enumeration trivial and shift the entire
security burden to server-side authorization, which this dataset shows
is frequently absent. Non-sequential UUIDs reduce predictability but do
not eliminate BOLA risk, as evidenced by 2.4\% of cases where UUIDs
were exploited through disclosure from other API responses. GraphQL's
\texttt{gid://service/Type/N} convention specifically encodes
sequential integers that become enumerable after base64 decoding;
platforms should not treat opacity of encoding as equivalent to access
control.

The Workflow-Context family (6.0\%) surfaces an anti-pattern where
authorization is evaluated at creation or access grant but not
re-evaluated when object state changes. The
deactivated-user/removed-staff pattern (Shopify \#2855610) and the
post-goes-private-but-timeline-still-returns-content pattern
(Automattic \#2258950) both reflect stale authorization enforcement
that standard RBAC frameworks do not automatically address.

\subsection{Implications for OWASP}

Table~\ref{tab:owasp} maps each taxonomy family against the attack
scenarios and prevention guidance in OWASP API1:2023, the canonical
reference for BOLA. The mapping reveals substantial variation in how
well OWASP API1:2023 represents empirically observed BOLA patterns.

\begin{table*}[htbp]
  \centering
  \caption{BOLA Taxonomy Families vs.\ OWASP API1:2023 Coverage}
  \label{tab:owasp}
  \small
  \begin{tabularx}{\textwidth}{p{1.6cm}p{1.2cm}p{2.6cm}p{3.0cm}X}
    \toprule
    \textbf{Family} & \textbf{Share} &
    \textbf{OWASP 2023 Scenario} &
    \textbf{OWASP Prevention} &
    \textbf{Coverage Assessment} \\
    \midrule
    Direct Object Reference BOLA
      & 36.9\%
      & Shop name enumeration to read revenue data
      & Enforce per-object authorization checks; prefer UUIDs over sequential IDs; write automated authorization tests
      & \textbf{Covered}: read-path identifier substitution is the paradigm case for OWASP's framing; UUID recommendation directly addresses this family \\
    \addlinespace
    Action-Level Object BOLA
      & 41.7\%
      & VIN substitution to control another user's vehicle; GraphQL \texttt{deleteReports} mutation without ownership validation
      & ``Enforce authorization checks for every resource access'' (generic); no explicit guidance on write/delete/trigger operations
      & \textbf{Partially covered}: the 2023 update added one delete-path example, but OWASP does not frame state-changing actions as a dominant distinct category or provide guidance that specifically targets write, delete, and trigger operations across ownership boundaries. The combined share of state-changing action types suggests that current OWASP guidance may underemphasize testing of state-changing ownership violations relative to their prevalence in observed disclosures \\
    \addlinespace
    Tenant Isolation BOLA
      & 8.3\%
      & None
      & None
      & \textbf{Partially covered}: Conceptually covered, but explicitly absent from examples, terminology, and prevention guidance \\
    \addlinespace
    Workflow-Context BOLA
      & 6.0\%
      & None
      & None
      & \textbf{Not covered}: OWASP guidance does not address authorization re-evaluation when object state changes (deactivation, archival, removal); static ownership checks at access-grant time are insufficient and this gap is not acknowledged \\
    \addlinespace
    Chained Disclosure BOLA
      & 4.8\%
      & Shop revenue scenario involves a prior list endpoint, loosely analogous
      & None
      & \textbf{Partially covered}: OWASP's shop scenario is structurally two-step but is not characterized as a distinct attack class; no guidance on identifier provenance, cross-endpoint leakage as an attack precondition, or multi-step authorization testing \\
    \addlinespace
    Object Rebinding BOLA
      & 2.4\%
      & None
      & None
      & \textbf{Not covered}: client-supplied ownership field manipulation (\texttt{owner\_id}, \texttt{account\_id}, \texttt{msg.Sender}) is absent from OWASP examples and prevention guidance entirely \\
    \addlinespace
    Vertical escalation
      & 11.9\%
      & Mentioned abstractly in 2019 framing; absent from 2023 examples
      & No vertical-specific testing recommendations
      & \textbf{Not explicitly represented}: all three OWASP 2023 examples are horizontal (peer-to-peer); no guidance distinguishes or recommends testing user-to-admin object access \\
    \bottomrule
  \end{tabularx}
\end{table*}

Three findings from this mapping have direct implications for OWASP
guidance.

First, Action-Level Object BOLA accounts for 41.7\% of confirmed cases
and appears alongside Direct Object Reference BOLA as one of the two
dominant families in the dataset. The OWASP framing has historically
emphasized identifier-substitution disclosure scenarios, while the 2023
revision introduced a single deletion example. However, it does not
frame state-changing actions as a dominant distinct category or provide
guidance that specifically targets write, delete, and trigger operations
across ownership boundaries. The combined share of state-changing
action types suggests that current OWASP guidance may underemphasize
testing of state-changing ownership violations relative to their
prevalence in observed disclosures.

Second, 11.9\% of confirmed cases are vertical (user-to-admin), yet
all three OWASP 2023 attack scenarios are horizontal. Prevention
guidance contains no vertical-specific recommendations. This is a
visibility gap: vertical BOLA carries disproportionately high severity
(Section~\ref{sec:severity}) and is absent from most testing playbooks
precisely because the canonical framing does not model it.

Third, Workflow-Context and Object Rebinding are absent from OWASP's
examples and prevention guidance, while Tenant Isolation is
conceptually encompassed by OWASP's object-level authorization model
but receives no explicit treatment. These collectively represent 16.7\%
of confirmed cases. The Workflow-Context family (6.0\%) is particularly
notable because it cannot be addressed by static ownership checks
alone: it requires re-evaluation of authorization when object lifecycle
state changes, an architectural requirement distinct from the
per-request checks OWASP recommends. The Action-Level Object family
additionally occupies the boundary zone between API1 (BOLA) and API5
(BFLA) in the OWASP Top 10: function access is granted, but
object-level authorization for state-changing actions is absent. This
boundary zone is where a significant share of real-world vulnerabilities
now lives.

\begin{figure}[htbp]
  \centering
  \includegraphics[width=\linewidth]{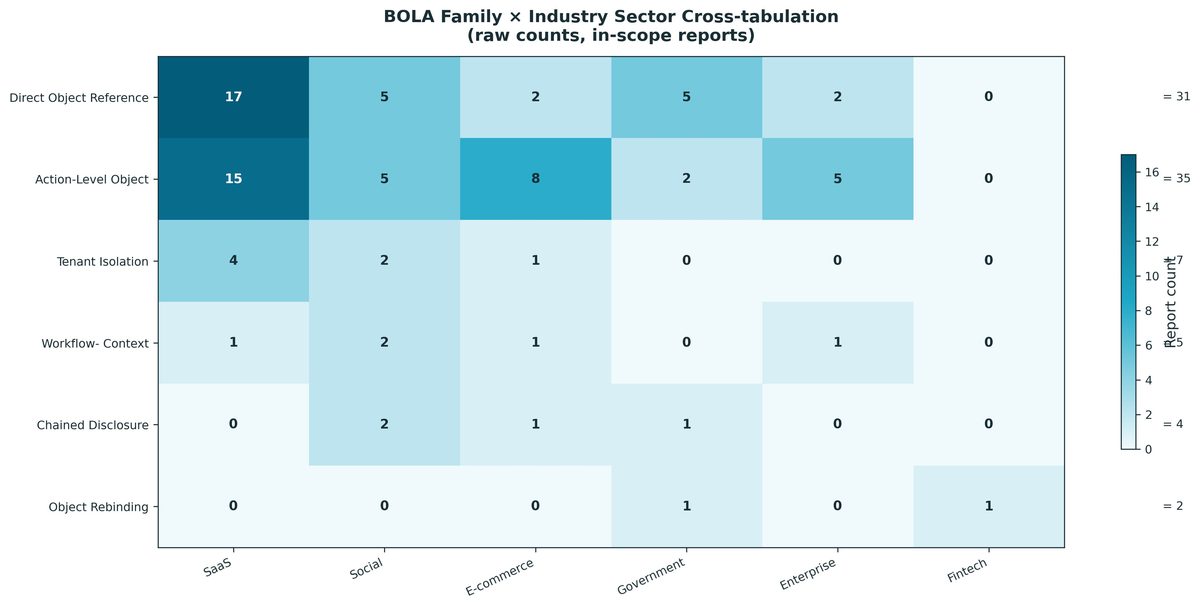}
  \caption{BOLA family $\times$ industry sector cross-tabulation
    (raw counts, $n = 84$). Action-Level Object BOLA dominates
    in E-commerce; Direct Object Reference leads in SaaS \&
    Productivity.}
  \label{fig:familysector}
\end{figure}

\section{Recommendations}

\subsection*{For Developers}
\begin{itemize}
  \item Derive object ownership from the authenticated session
    server-side on every request. Never read ownership from the
    request body or query parameter.
  \item Apply ownership scoping at the database query layer, not in
    application code after retrieval.
  \item Re-evaluate authorization when object state changes.
    Deactivation, archival, and removal must invalidate existing
    access grants to that object.
  \item Treat all identifier formats as potentially enumerable. UUIDs
    and encoded IDs reduce predictability but do not eliminate BOLA
    risk if object identifiers are leaked elsewhere in your API.
  \item Avoid GraphQL Global IDs that encode sequential backend
    integers. If you use them, ensure the authorization check occurs
    at the resolver level, not just at the schema level.
\end{itemize}

\subsection*{For Testers}
\begin{itemize}
  \item Test all HTTP actions, not just \texttt{GET}. Action-Level
    Object BOLA is now the single largest confirmed family; read-only
    test protocols miss the dominant real-world pattern.
  \item Provision two accounts with distinct object ownership before
    testing. Single-account testing cannot confirm a cross-user
    boundary failure.
  \item Test objects in all lifecycle states: active, archived,
    deactivated, deleted. Authorization failures on state-changed
    objects are a distinct family that standard test flows miss.
  \item Map identifier-returning endpoints before testing
    authorization. Any endpoint that returns another user's object
    identifier is a potential chained disclosure source.
  \item Include tenant-boundary testing for any multi-tenant SaaS
    application. Register two separate organizations and verify that
    requests from one cannot access the other's objects.
  \item Include vertical BOLA in test scope: test whether a
    lower-privileged session can act on objects owned by
    higher-privileged roles. 11.9\% of confirmed cases are vertical,
    and they are absent from most testing playbooks.
\end{itemize}

\section{Limitations}
\label{sec:limitations}

\paragraph{Selection bias via HackerOne coverage.}
The sampling frame is limited to programs that operate on HackerOne,
accept BOLA-type reports, and publicly disclose findings. The dataset
over-represents developer tooling platforms (GitHub, GitLab, HackerOne
itself) and under-represents fintech, healthcare, and
enterprise-internal APIs.

\paragraph{Disclosure depth heterogeneity.}
The 21 Low-confidence classifications (25.0\%) reflect cases where
cross-user access was implied but not demonstrated. Excluding them
yields 63 High+Medium confidence reports with higher internal validity.

\paragraph{Classifier imprecision.}
LLM-assisted classification introduces systematic errors that manual
review partially mitigates. The Direct Object Reference vs.\
Action-Level Object distinction requires a judgment about whether the
primary action is state-changing; this is ambiguous when reports
describe both reading and modifying.

\paragraph{Temporal confounding.}
36 reports with undetermined disclosure year preclude robust trend
analysis. Apparent stability across 2023--2026 may reflect true
persistence or a disclosure lag.

\paragraph{Single-program concentration.}
HackerOne (13 reports, 15.5\%) and U.S.\ Department of Defense (9
reports, 10.7\%) together contribute 26.2\% of the in-scope dataset
across 29 programs, 17 of which contribute a single report.
Program-weighted reanalysis (Section~\ref{sec:robustness}) quantifies
the effect. Findings robust to this reweighting, Direct Object
Reference prevalence, SaaS \& Productivity sector dominance, and the
combined 70\% share of the two leading families, can be stated with
higher confidence.

\section{Conclusion}

We present one of the first systematic, taxonomy-driven empirical
analyses of BOLA in real-world bug bounty disclosures. From 200
HackerOne candidates, we confirmed 84 in-scope instances across six
families. Key takeaways for practitioners:

\begin{itemize}
  \item Action-Level Object BOLA (41.7\%) and Direct Object Reference
    BOLA (36.9\%) emerge as the two dominant observed families,
    jointly accounting for 78.6\% of confirmed cases.
  \item Sequential integers remain the most evidenced known identifier
    type (36.9\%) in 2023--2026 disclosures from mature security
    programs.
  \item Non-sequential identifiers do not eliminate BOLA risk. Encoded
    ID, username, email, and UUID-based cases account for 39.2\% of
    known-format reports.
  \item 11.9\% of BOLA is vertical (user-to-admin), a pattern absent
    from most developer guidance.
  \item Only 84 of 200 sampled IDOR/IAC-tagged disclosures (42.0\%)
    were ultimately confirmed as in-scope BOLA, highlighting
    substantial divergence between platform tagging and rigorous
    classification.
  \item GraphQL global IDs are systematically exploited through
    decode/increment/re-encode patterns across HackerOne, GitLab, and
    Shopify.
  \item Critical-severity cases (10.7\%) are concentrated in
    government-sector reports and fintech platform-level failures.
\end{itemize}

Future work should extend the sampling frame to closed-source programs,
examine BOLA patterns in OpenAPI specifications to characterize
preventable root causes, and evaluate automated detection coverage
against the six-family taxonomy as a benchmark.

\section*{Acknowledgements}

The author thanks Jose Haro Peralta for his supervision, guidance, and
for feedback on earlier drafts of this manuscript. This work was
conducted under the auspices of APIsec Research Labs, whose support
made this work possible.

\section*{Ethics Statement}

All analyzed vulnerabilities were drawn exclusively from publicly
disclosed HackerOne reports that had already been remediated and
released by the affected programs. No active exploitation, interaction
with non-public systems, or collection of private user data was
performed as part of this study.

\bibliographystyle{plainnat}
\bibliography{bola_refs}

\appendix

\section{Classification Prompt (Abbreviated)}
\label{app:prompt}

The full classifier prompt (\texttt{prompts/classification.md}) is
available in the supplementary repository
(\url{https://github.com/hackwither/bola-in-the-wild}). Key components
include explicit BOLA vs.\ BFLA disambiguation rules; operational
confidence level definitions (High: endpoint path OR request example
AND confirmed cross-user access; Medium: cross-user access evident but
one technical detail missing; Low: cross-user access implied but
unconfirmed); endpoint normalization rules; and six-family definitions
with distinguishing criteria for adjacent families.

\section{Sampling Frame Reproducibility}
\label{app:sampling}

The sampling frame was queried via the HackerOne GraphQL API
(\url{https://hackerone.com/graphql}) with weakness names ``IDOR,''
``Improper Access Control,'' and ``Broken Access Control,'' disclosed
2021-01-01 to 2026-01-31, sorted by upvote count descending, first 200
results. The query and paginated output are preserved in the
supplementary repository.

\section{Family Distribution by Industry Sector}
\label{app:crosstab}

\begin{table}[htbp]
  \centering
  \caption{BOLA Family Distribution by Industry Sector
    (raw counts, $n = 84$)}
  \label{tab:sectorcrosstab}
  \small
  \begin{tabular}{lrrrrrrr}
    \toprule
    \textbf{Sector} &
    \makecell{\textbf{Dir.}\\\textbf{OR}} &
    \makecell{\textbf{Act.}\\\textbf{Lvl}} &
    \makecell{\textbf{Ten.}\\\textbf{Isol.}} &
    \textbf{Wflow} &
    \textbf{Chain} &
    \makecell{\textbf{Obj.}\\\textbf{Reb.}} &
    \textbf{Tot.} \\
    \midrule
    SaaS \& Prod.     & 17 & 15 & 4 & 1 & 0 & 0 & 37 \\
    Social \& Con.    &  5 &  5 & 2 & 2 & 2 & 0 & 16 \\
    E-com. \& Mkt.    &  2 &  8 & 1 & 1 & 1 & 0 & 13 \\
    Gov. \& Public    &  5 &  2 & 0 & 0 & 1 & 1 &  9 \\
    Ent. \& Infra.    &  2 &  5 & 0 & 1 & 0 & 0 &  8 \\
    Fintech           &  0 &  0 & 0 & 0 & 0 & 1 &  1 \\
    \midrule
    Total             & 31 & 35 & 7 & 5 & 4 & 2 & 84 \\
    \bottomrule
  \end{tabular}
\end{table}

Key patterns:
\begin{itemize}
  \item Action-Level Object BOLA dominates in E-commerce (8 of 13
    reports, 61.5\%), driven by ride-hailing bid manipulation, driver
    rating fraud, and forced trip acceptance, all write operations on
    another user's transaction objects.
  \item In SaaS \& Productivity, Direct Object Reference (17) and
    Action-Level Object (15) are nearly split, suggesting mature
    platforms have partially addressed read-path BOLA but less so
    write-path.
  \item Chained Disclosure clusters in Social \& Consumer Platforms
    (2 of 4 instances) alongside E-commerce (1), consistent with
    multi-step transaction and feed workflows.
  \item Object Rebinding is confined to Government (DoD Salesforce
    cloneAttachment) and Fintech (Cosmos SDK), reflecting cases where
    server-trusted client-supplied ownership fields appear in complex
    message-passing or document-attachment architectures.
  \item Tenant Isolation is entirely absent from Government, Enterprise,
    and Fintech in this sample, concentrated in SaaS (4), Social (2),
    and E-commerce (1), where multi-tenancy is an explicit
    architectural feature.
\end{itemize}

\end{document}